\documentclass[aps,prb,twocolumn,a4paper,10pt,superscriptaddress,longbibliography]{revtex4-1}

\usepackage{graphicx}
\usepackage{amsmath}
\usepackage{amssymb}
\usepackage{amsfonts}
\usepackage{dsfont}
\usepackage{braket}
\usepackage{color}
\usepackage{braket,slashed}
\usepackage[mathscr]{euscript}
\usepackage{hyperref}
\usepackage{subfigure}
\usepackage{xfrac}
\usepackage{bm}
\usepackage{kantlipsum}
\usepackage{microtype}
\usepackage{fancyhdr}

\allowdisplaybreaks[4]

\renewcommand{\dag}{^\dagger}

\newcommand{\e}{\ensuremath{\mathrm{e}}}

\newcommand{\diagram}[1]{\vcenter{\hbox{\includegraphics[scale=0.6]{./#1.pdf}}}}
\newcommand{\diagramm}[1]{\vcenter{\hbox{\includegraphics[scale=0.4]{./#1.pdf}}}}
\newcommand{\diagrammm}[1]{\vcenter{\hbox{\includegraphics[scale=0.5]{./#1.pdf}}}}

\begin{document}

\title{Gradient methods for variational optimization of projected entangled-pair states}
\author{Laurens Vanderstraeten}
\author{Jutho Haegeman}
\affiliation{Ghent University, Department of Physics and Astronomy, Krijgslaan 281-S9, B-9000 Gent, Belgium}
\author{Philippe Corboz}
\affiliation{Institute for Theoretical Physics, University of Amsterdam, Science Park 904 Postbus 94485, 1090 GL Amsterdam, The Netherlands}
\author{Frank Verstraete}
\affiliation{Ghent University, Department of Physics and Astronomy, Krijgslaan 281-S9, B-9000 Gent, Belgium}
\affiliation{Vienna Center for Quantum Science, Universit\"at Wien, Boltzmanngasse 5, A-1090 Wien, Austria}

\date{\today}

\begin{abstract}
We present a conjugate-gradient method for the ground-state optimization of projected entangled-pair states (PEPS) in the thermodynamic limit, as a direct implementation of the variational principle within the PEPS manifold. Our optimization is based on an efficient and accurate evaluation of the gradient of the global energy functional by using effective corner environments, and is robust with respect to the initial starting points. It has the additional advantage that physical and virtual symmetries can be straightforwardly implemented. We provide the tools to compute static structure factors directly in momentum space, as well as the variance of the Hamiltonian. We benchmark our method on Ising and Heisenberg models, and show a significant improvement on the energies and order parameters as compared to algorithms based on imaginary-time evolution.
\end{abstract}

\maketitle


\section{Introduction}

Ever since the birth of quantum mechanics, the quantum many-body problem has been at the center of theoretical and computational physics. Despite the simplicity of the fundamental equations, it has been notoriously difficult to simulate the quantum behavior of many-body systems. This is especially true for low-dimensional systems: because quantum correlations are stronger, perturbation theory often fails and more sophisticated methods are needed. This cry for better methods is loudest with respect to two-dimensional systems, because of the large range of unexplored quantum phenomena---quantum spin liquids \cite{Balents2010}, topological order \cite{Zeng2015} and quasi-particle fractionalization  are only three examples.
\par Because Monte Carlo sampling is often plagued by the sign problem and exact diagonalization is necessarily limited to small system sizes, it seems that variational methods are the way to go for exploring the two-dimensional quantum world. There are essentially two prerequisites for a successful variational approach: (i) an adequate variational ansatz that captures the physics for the problem at hand, and (ii) an efficient way of computing observables and optimizing the variational parameters. Examples such as the density-matrix renormalization group \cite{Stoudenmire2012,Yan2011} and Gutzwiller-projected wave functions \cite{Sorella2005,Iqbal2013} seem to meet the latter, but it is unclear to what extent they are the natural choice for simulating two-dimensional quantum systems.
\par In recent years projected entangled-pair states (PEPS) \cite{Verstraete2004b, Orus2013} have emerged as a viable candidate for capturing the physics of ground states of strongly-correlated quantum lattice models in two dimensions. It is by explicitly modeling the distribution of entanglement in low-energy states of local Hamiltonians, that PEPS parametrize the ``physical corner of Hilbert space''. Indeed, PEPS have a built-in area law for the entanglement entropy \cite{Eisert2010a}, they provide a natural characterization of topological order \cite{Schuch2010a, Wahl2013a, Sahinoglu2014, Buerschaper2014, Bultinck2015}, and they can realize bulk-boundary correspondences explicitly \cite{Cirac2011, Schuch2013a, Yang2014}. Moreover, PEPS can be formulated directly in the thermodynamic limit \cite{Jordan2008} which allows us to focus on bulk physics without any finite-size or boundary effects.
\par An efficient optimization of the parameters in a PEPS has proven to be more challenging. According to the variational principle, finding the best approximation to the ground state for a given Hamiltonian $H$ reduces to the minimization of the energy expectation value. For infinite PEPS this amounts to a highly non-linear optimization problem for which the evaluation of, e.g., the gradient of the energy functional is a hard problem. For that reason, the state-of-the-art PEPS algorithms have taken recourse to imaginary-time evolution \cite{Jordan2008, Jiang2008, Phien2015}: a trial PEPS state is evolved with the operator $\e^{-\tau H}$, which should result in a ground-state projection for very long times $\tau$. This imaginary-time evolution is integrated by applying small time steps $\delta\tau$ with a Trotter-Suzuki decomposition and, after each time step, truncating the PEPS bond dimension in an approximate way. This truncation can be done by a purely local singular-value decomposition---the so-called simple-update\cite{Jiang2008} algorithm---or by taking the full PEPS wave function into account---the full-update \cite{Jordan2008} or fast full-update \cite{Phien2015} algorithm. 
\par These imaginary-time algorithms have allowed very accurate simulations of frustrated spin systems \cite{Corboz2012, Zhao2012, Gu2013, Corboz2013, Iregui2014, Corboz2014a, Picot2015, Picot2015a, Nataf2016, Corboz2016} and strongly correlated electrons \cite{Corboz2010a, Corboz2010b, Corboz2014, Corboz2016}, but it remains unclear whether they succeed in finding the optimal state in a given variational class of PEPS. Although computationally very cheap, ignoring the environment in the simple-update scheme is often a bad approximation for systems with large correlations. The full-update scheme takes the full wave function into account for the truncation, but requires the inversion of the effective environment which is potentially badly conditioned. This problem was solved by regularizing the environment appropriately and fixing the gauge of the PEPS tensor \cite{Lubasch2014, Phien2015}. Nonetheless, the truncation procedure in the full-update scheme is not guaranteed to provide the \emph{globally} optimal truncated tensors in the sense that the global overlap of the truncated and the original PEPS is maximized. Indeed, the truncated tensor is optimized \emph{locally} and afterwards put in at every site in the lattice to give an updated (global) PEPS wave function.
\par Similar issues have been at the center of attention in the context of matrix product states (MPS) \cite{Schollwock2011a}, the one-dimensional counterparts of PEPS, where a number of different strategies have been around for optimizing ground-state approximations directly in the thermodynamic limit \cite{Vidal2007b, McCulloch2008, Haegeman2011d}. Recently, the problem of finding an optimal matrix product state has been reinterpreted by (i) identifying the class of matrix product states as a non-linear manifold embedded in physical Hilbert space \cite{Haegeman2013b, Haegeman2014}, and (ii) formulating a minimization problem of the global energy functional on this manifold. A globally optimal state can then be recognized as a point on the manifold for which the gradient of the energy functional is zero. Moreover, approximating time evolution within the manifold is optimized, as dictated by the time-dependent variational principle \cite{Haegeman2011d}, by projecting the time evolution onto the tangent space of the manifold. In the case of imaginary time, this tangent vector is exactly the gradient, which shows that different optimization algorithms can be compared within this unifying manifold interpretation \cite{Haegeman2014a}. Moreover, whereas imaginary-time evolution more or less corresponds to a steepest-descent method \cite{Haegeman2013b}, more advanced optimization methods such as conjugate-gradient or quasi-Newton algorithms can find an optimal matrix product state much more efficiently \cite{Milsted2013, prepJutho}.
\par In Ref.~\onlinecite{Vanderstraeten2015b} it was shown how to implement these \emph{tangent space methods} for PEPS by introducing a contraction scheme based on the concept of a ``corner environment''. Building on that work, this paper presents a PEPS algorithm that optimizes the global energy functional using a conjugate-gradient optimization method. In contrast to other methods, this algorithm has a clear convergence criterion, which can guarantee that an optimal state has been reached. Moreover, it allows us to more easily impose physical symmetries on the PEPS. On the fly, the contraction scheme also allows us to compute the energy variance of the variational ground state---an unbiased measure of the accuracy of the variational ansatz \textit{and} a tool for better energy extrapolations---as well as general two-point correlation functions and static structure factors.
\par In the next section [Sec.~\ref{sec:env}] we review the ``corner environment'' in considerable detail and show how to compute static structure factors of a PEPS. Next [Sec.~\ref{sec:conj}] we discuss our conjugate-gradient scheme for the PEPS optimization, and explain how to evaluate the energy gradient and the energy variance. We benchmark [Sec.~\ref{sec:bench}] our method by applying it to the transverse Ising model, the XY model and the isotropic Heisenberg model. In the last section [Sec.~\ref{sec:concl}] we discuss the possible extensions and applications.


\section{Effective environments and two-point correlation functions}
\label{sec:env}

Consider an infinite square lattice with every site hosting a quantum degree of freedom with dimension $d$. For this quantum spin system, a PEPS can be introduced formally as
\begin{equation} \label{peps}
\ket{\Psi(A)} = \sum_{\{s\}} \mathcal{C}_2(A) \ket{\{s\}}
\end{equation}
where $\mathcal{C}_2(\dots)$ is the contraction of an infinite tensor network. This contraction is most easily represented graphically as
\begin{equation*} 
\mathcal{C}_2(A) = \diagram{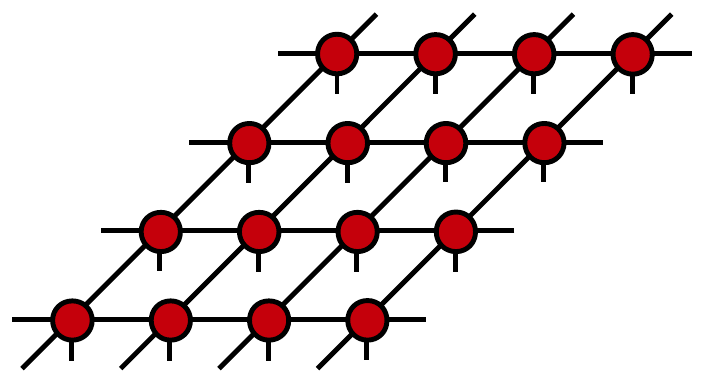},
\end{equation*}
with the red circle always representing the same five-legged tensor $A$,
\begin{equation*}
A_{u,r,d,l}^s = \diagram{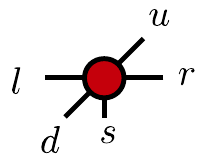}.
\end{equation*}
In order to obtain a physical state, a tensor $A$ is associated with every site in the lattice and all virtual indices $(u,r,d,l)$ are contracted in the network. The physical indices $s$ are left open, such that a coefficient is obtained for every spin configuration in the superposition in Eq.~\eqref{peps}. The graphical representation is then obtained by connecting links that are contracted and leaving the physical links open. The virtual degrees of freedom in the PEPS carry the quantum correlations and mimic the entanglement structure of low-energy states. The dimension of the virtual indices is called the bond dimension $D$ and can be tuned in order to enlarge the variational class; as such, it acts as a refinement parameter for the variational PEPS ansatz.
\par The norm of an infinite PEPS can be pictorially represented as
\begin{equation*}
\braket{\Psi(A)|\Psi(A)} = \diagrammm{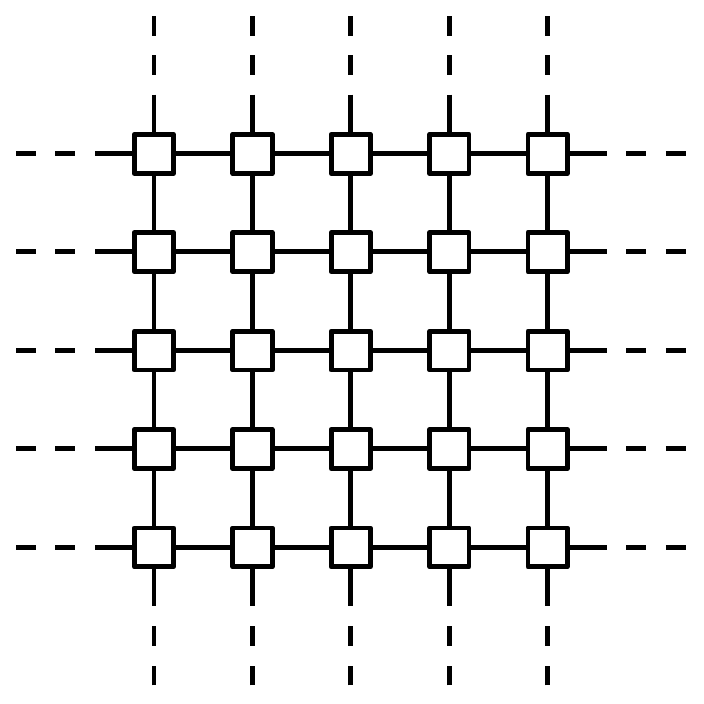},
\end{equation*}
where every block represents the tensor $a$ obtained by contracting the tensor $A$ with its conjugate $\bar{A}$ over the physical index, i.e.,
\begin{equation*}
a = \diagrammm{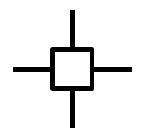} = \diagrammm{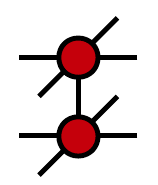}
\end{equation*}
As in the rest of this paper, the virtual indices of the ket and bra level are grouped into one index, so that these ``top view'' representations of double-layer tensor contractions are simplified.
\par The norm of the PEPS is thus obtained by the contraction of an infinite tensor network and can, in general, only be done approximately. Different numerical methods have been developed to contract these infinite networks efficiently, which allows the evaluation of the norm of a PEPS, as well as expectation values and correlation functions.

\subsection{The linear transfer matrix}

The first and most straightforward strategy is based on the linear transfer matrix $\mathcal{T}$, graphically represented as
\begin{equation*}
\mathcal{T} = \diagrammm{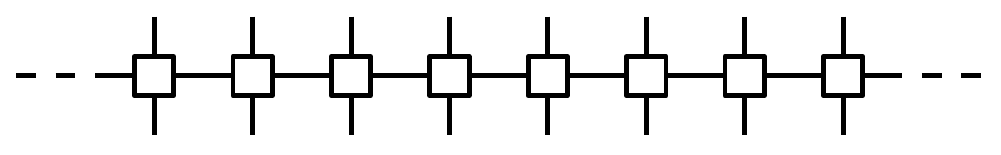}.
\end{equation*}
This object carries all the correlations in the PEPS from one row in the network to the next. One can of course define a similar transfer matrix in the vertical direction, and even diagonal transfer matrices can be considered. Naturally the transfer matrix is interpreted as an operator from the top to the bottom indices, so that the full contraction of the two-dimensional network reduces to successively multiplying copies of $\mathcal{T}$. In the thermodynamic limit, the norm of a PEPS is thus given by
\begin{equation*}
\braket{\Psi(A)|\Psi(A)} = \lim_{N\rightarrow\infty} \mathcal{T}^N = \lambda^N
\end{equation*}
with $\lambda$ the leading eigenvalue of the transfer matrix. The associated leading eigenvector or fixed point contains all the information on the correlations of a half-infinite part of the lattice. 
\par An exact representation of the fixed point is only possible in a number of special cases and approximate methods have to be devised in general. Given the versatility of matrix product states (MPS) for approximating the ground state of local gapped Hamiltonians \cite{Schollwock2011a}, one expects that this class of states might provide a good variational ansatz for the case of gapped transfer matrices as well. Moreover, the bond dimension of the matrix product state representation of the fixed point, denoted with $\chi$, can be tuned systematically, such that the errors can be kept under control perfectly. Whereas MPS approximations for fixed points go way back \cite{Baxter1968}, a variety of efficient tensor-network methods have been developed \cite{Verstraete2004b, Orus2008} recently. Here we use an algorithm \cite{prepJutho} in the spirit of Ref.~\onlinecite{Haegeman2014a}, which treats the linear and corner transfer matrices [Sec.~\ref{sec:ctm}] on a similar footing.
\par The fixed-point equation can be stated graphically as
\begin{multline} \label{linear}
\diagrammm{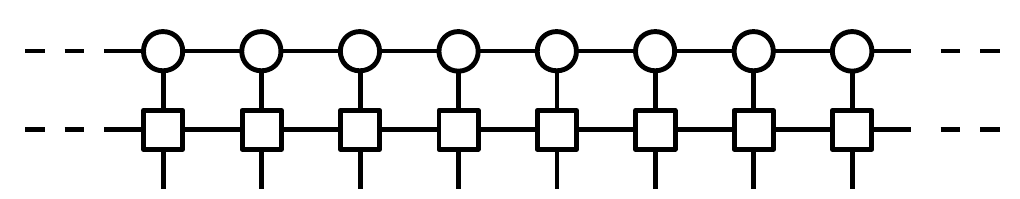} \\ \approx \lambda \diagrammm{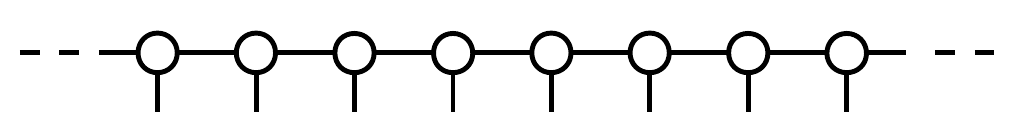}.
\end{multline}
For this equation to hold, a relation of the form
\begin{equation*}
\diagrammm{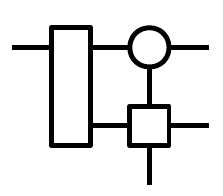} \approx \diagrammm{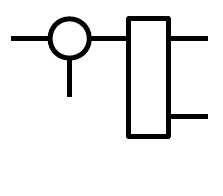}
\end{equation*}
should hold to a very high precision \cite{prepJutho}. Indeed, if this tensor (rectangle) exists, it maps the action of the transfer matrix back to the same MPS fixed point. The virtual dimension of the MPS fixed point will be denoted as $\chi$ and can be tuned to improve the accuracy of the PEPS contraction.
\par Given that the fixed-point equation can be solved efficiently, the PEPS can now be normalized to one by rescaling the $A$ tensor such that the largest eigenvalue $\lambda$ of the transfer matrix equals unity. With the MPS fixed point, the expectation value of a local operator at an arbitrary site $i$,
\begin{equation*}
\braket{O_i} = \frac{\bra{\Psi(A)} O_i \ket{\Psi(A)}} {\braket{\Psi(A)|\Psi(A)}},
\end{equation*}
can be easily computed. First the upper and lower halves of the network are replaced by the fixed points,
\begin{multline*}
\diagrammm{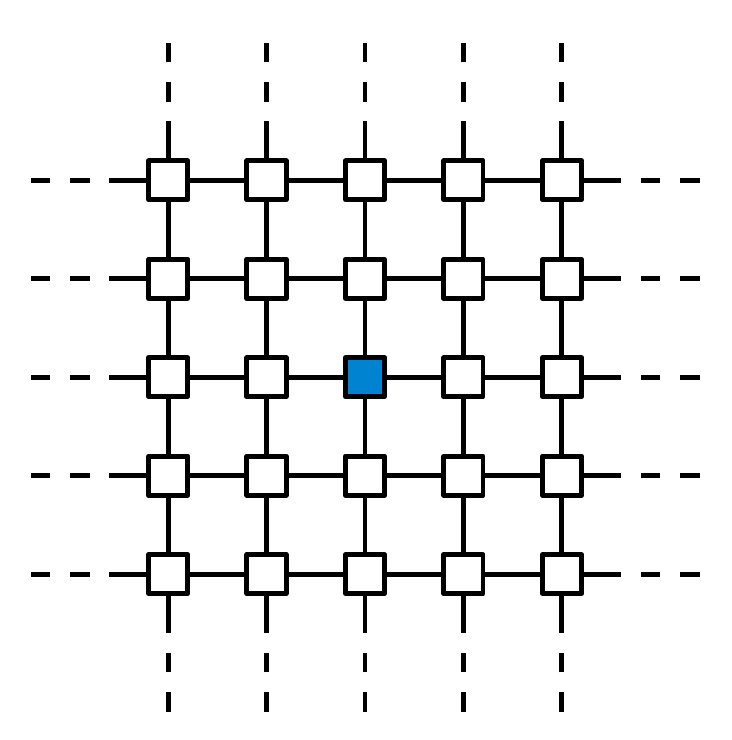} \\ \approx \diagrammm{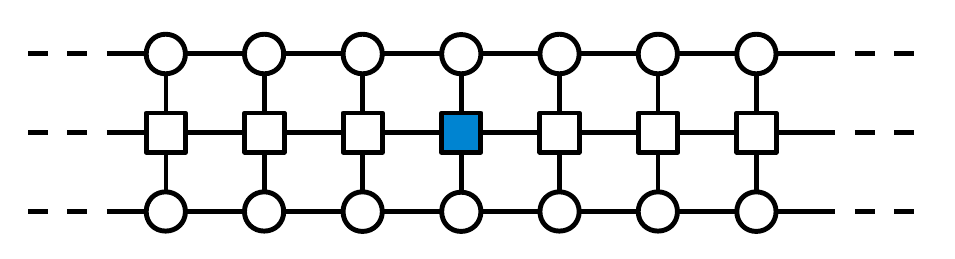},
\end{multline*}
where a colored block tensor always indicates the presence of a physical operator at that site. The resulting effective one-dimensional network can be evaluated exactly by finding the leading left and right eigenvectors (fixed points) of the channel operator, 
\begin{equation*}
\diagrammm{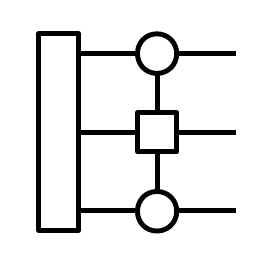} = \mu \diagrammm{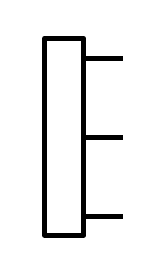}
\end{equation*}
and 
\begin{equation*}
\diagrammm{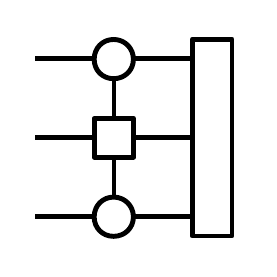} = \mu \diagrammm{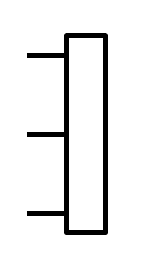}.
\end{equation*}
The eigenvalue $\mu$ depends on the normalization of the MPS tensors in the upper and lower fixed points of the linear transfer matrix, and its value can be put to one. The fixed points are determined up to a factor, which can be fixed by imposing that the norm of the PEPS
\begin{equation*}
\braket{\Psi(A)|\Psi(A)} = \diagrammm{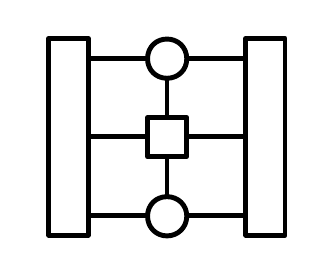}
\end{equation*}
equals unity such that 
\begin{equation*}
\braket{O_i} = \bra{\Psi(A)} O_i \ket{\Psi(A)} = \diagrammm{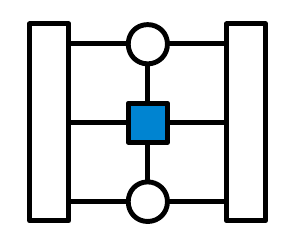}.
\end{equation*}

\subsection{The corner transfer matrix}
\label{sec:ctm}

Another set of methods for contracting two-dimensional tensor networks relies on the concept of the corner transfer matrix, which was first applied to classical lattice systems \cite{Baxter1968, Baxter1981, Nishino1996, Nishino1997} and recently used extensively in tensor network simulations \cite{Orus2009, Corboz2010b, Orus2012}. The strategy now is to break up the infinite tensor network in different regions, and represent these as tensors with a fixed dimension. Graphically, the set up is
\begin{equation*}
\diagrammm{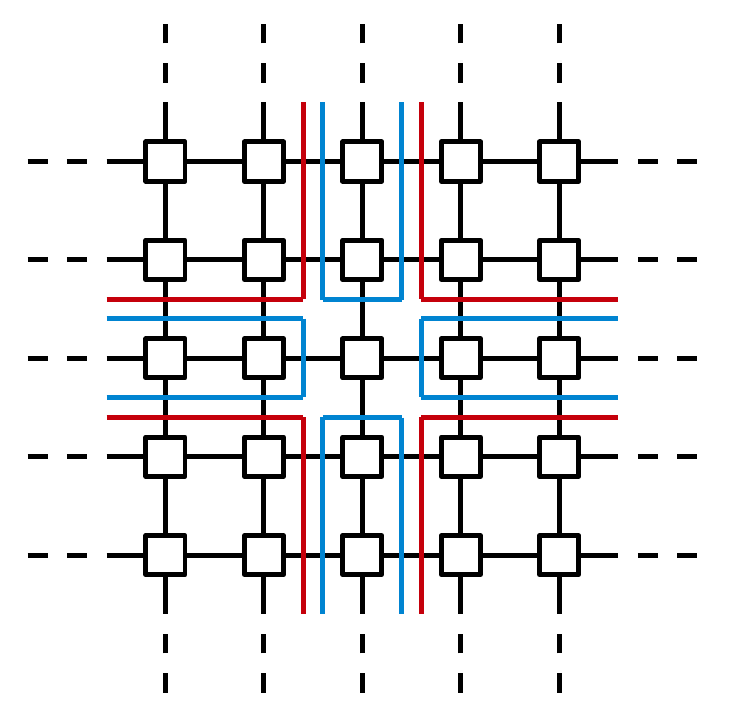} \approx \diagrammm{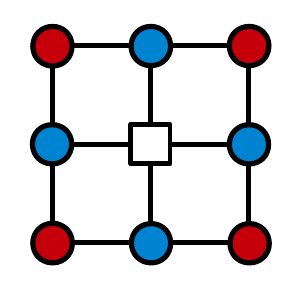}.
\end{equation*}
A red tensor represents the compression of one of the corners of the network, whereas the blue tensors capture the effect of an infinite row of $a$ tensors. Together, they provide an effective one-site environment for the computation of the norm of the PEPS or local expectation values.
\par This scheme can now be extended \cite{Vanderstraeten2015b} in order to evaluate non local expectation values such as general two-point correlation functions. Indeed, by not compressing the blue region above, one can construct an environment that looks like
\begin{equation*}
\diagrammm{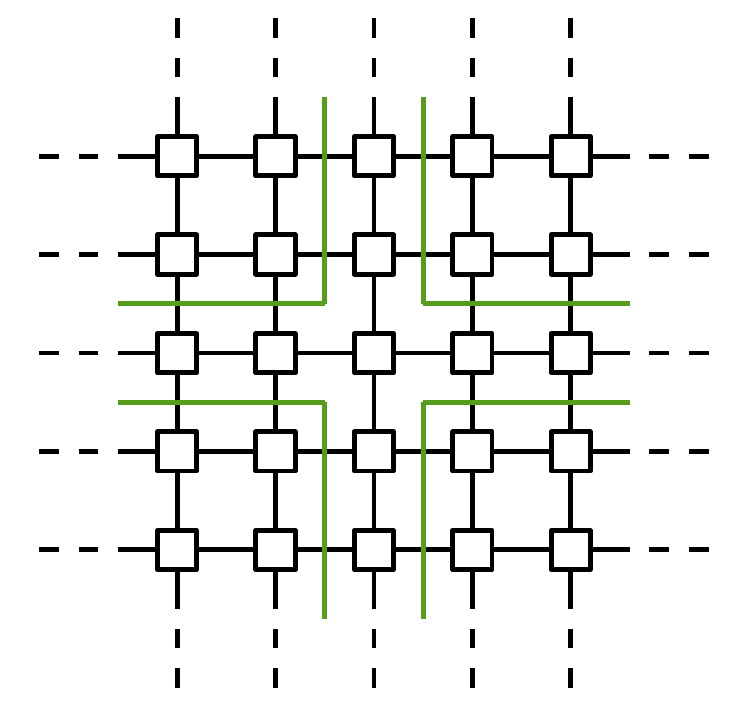}  \approx \diagrammm{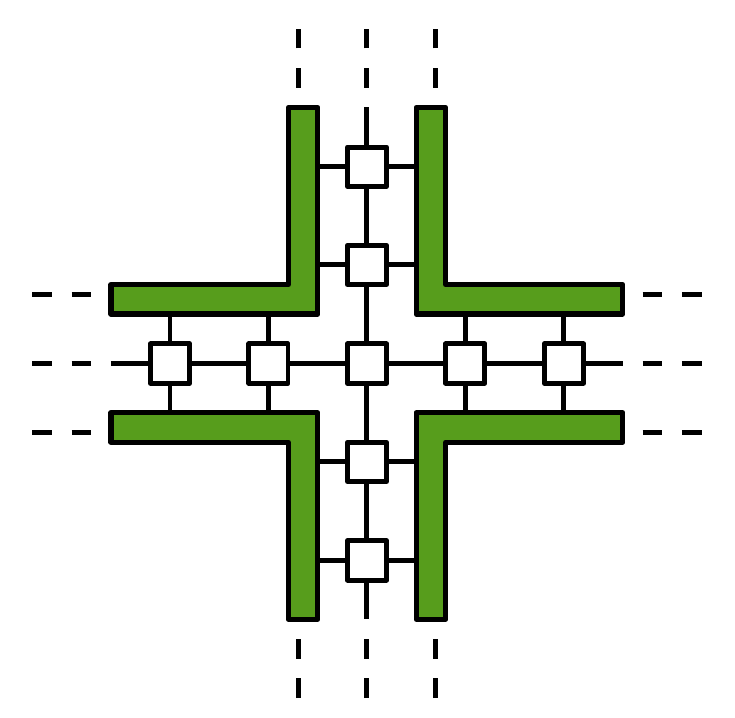},
\end{equation*}
so that one could evaluate operators that have an arbitrary location in the lattice.
\par Finding this effective ``corner environment'' can again be done by solving a fixed-point equation. Indeed, the green corner-shaped environment should be the result of an infinite number of iterations of an equally corner-shaped transfer matrix; the fixed-point equation is
\begin{equation*}
\diagrammm{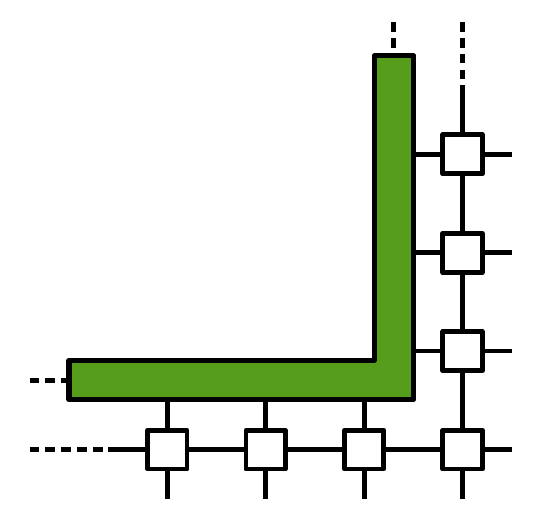} \propto \diagrammm{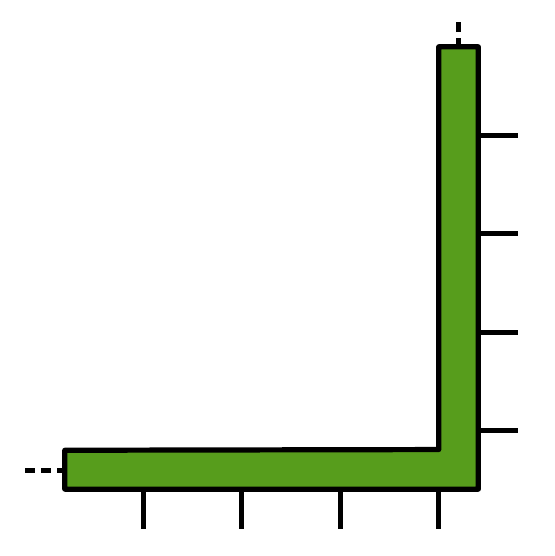}.
\end{equation*}
Very far from the corner this equation reduces to the one for the linear transfer matrix. This implies that, asymptotically, the fixed point can be well approximated by an MPS. Let us therefore make the ansatz that the full fixed point can be approximated as an MPS, where we put an extra tensor on the virtual level to account for the corner. With this ansatz, the fixed point equation is given by
\begin{equation} \label{corner}
\diagrammm{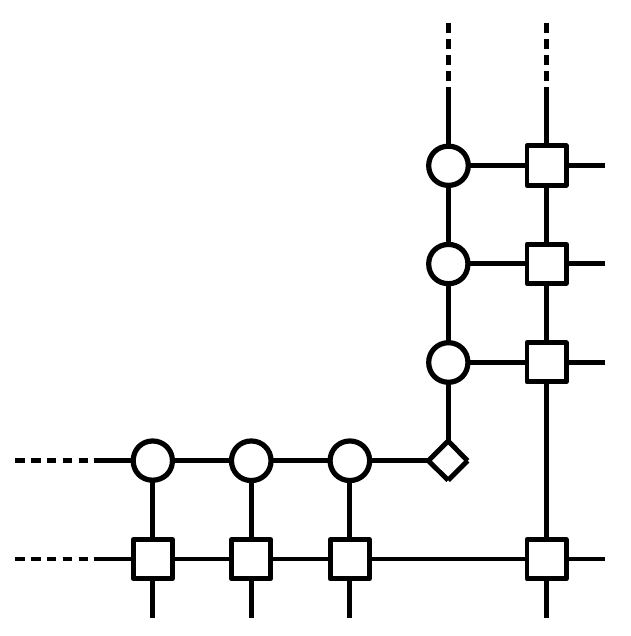} \propto \diagrammm{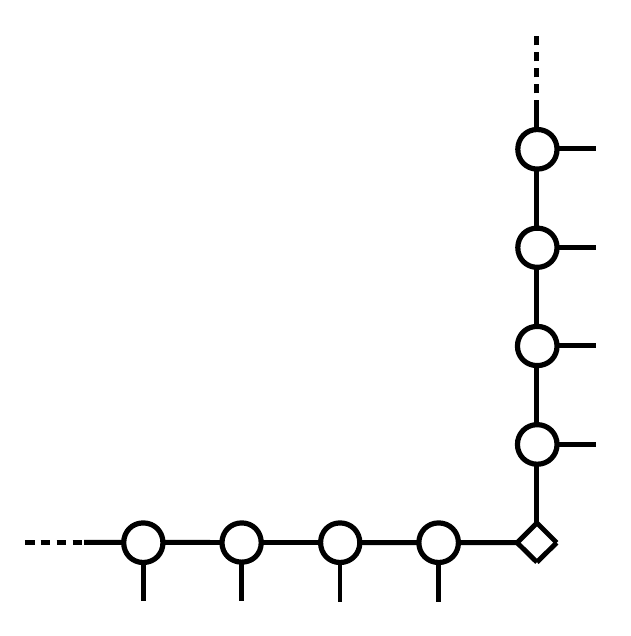}.
\end{equation}
We expect \cite{Vanderstraeten2015b} that this fixed point can be modelled using the MPS tensors from the fixed points of the linear transfer matrix, up to the corner matrix, which captures the effect of the corner shape. With this ansatz, we obtain a linear fixed point equation for the corner matrix, which corresponds to a simple eigenvalue equation and can be solved efficiently.

\subsection{Channel environments}

Once we have found (i) the fixed points of the linear transfer matrix in all directions [Eq.~\eqref{linear}], and (ii) the four corner tensors [Eq.~\eqref{corner}], we can contract the network corresponding to the norm, a local expectation value, or a correlation function of the PEPS. Let us assume that the tensor $A$ is normalized such that the largest eigenvalue of the linear transfer matrix is unity. Computing the norm $\braket{\Psi(A)|\Psi(A)}$ with a channel environment then reduces to the contraction of
\begin{equation*}
\diagramm{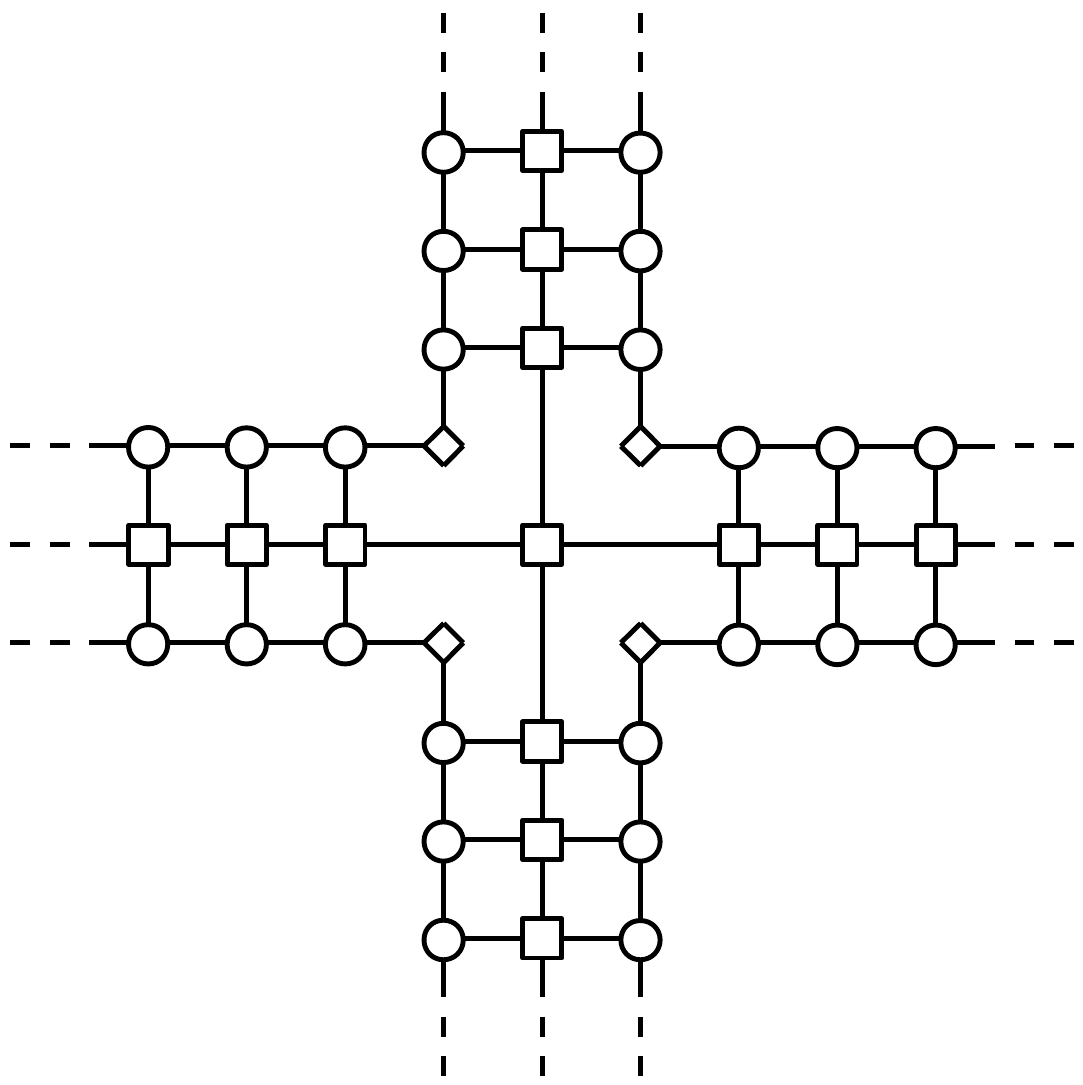}.
\end{equation*}
An infinitely long channel can be contracted by computing the fixed point $\rho_L$ of the ``channel operator''. Therefore the eigenvector corresponding to the largest eigenvalue should be found, i.e.,
\begin{equation*}
\diagramm{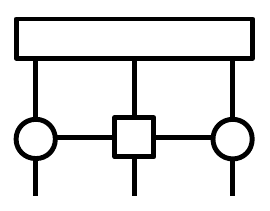} = \lambda \times \diagramm{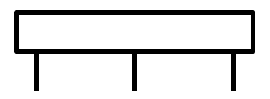}
\end{equation*}
for the top channel. The boundary MPS tensors have to be rescaled such that the largest eigenvalue $\lambda$ is put to one. Similarly, the fixed point in the other direction $\rho_R$ is defined as
\begin{equation*}
\diagramm{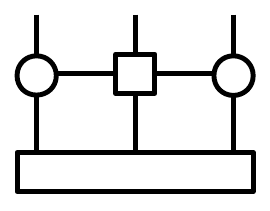} = \lambda \times \diagramm{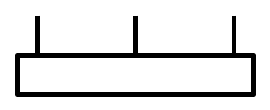}
\end{equation*}
The inner product of the left and right fixed points is put to one. For further use, we note that, by subtracting the projector on the largest eigenvector, an operator is constructed that has spectral radius strictly smaller than one,
\begin{equation*}
\rho\left( \diagramm{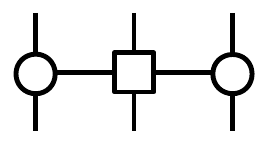} - \diagramm{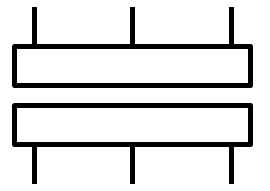} \right) < 1.
\end{equation*}
The norm of the PEPS is then reduced to
\begin{equation*}
\braket{\Psi(A)|\Psi(A)} = \diagramm{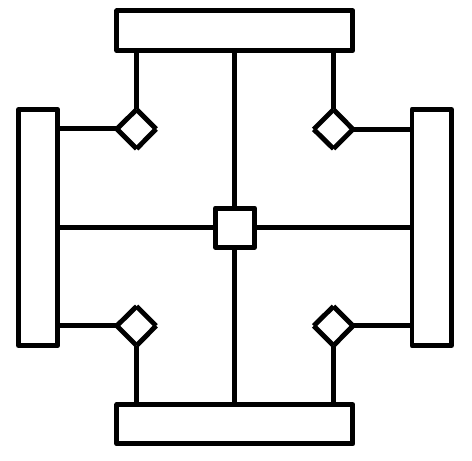},
\end{equation*}
which can be scaled to 1 by rescaling the corner tensors by the appropriate scalar. With these conventions, the norm of the state is well defined and expectation values can be safely computed. For a local one-site operator $O$ we have
\begin{equation*}
\bra{\Psi(A)}O\ket{\Psi(A)} = \diagramm{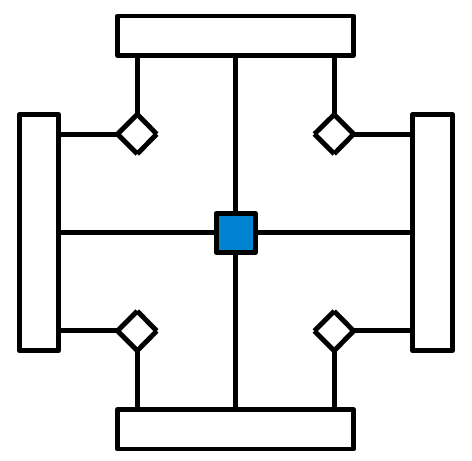},
\end{equation*}
and, similarly, the expectation value of a two-site operator is
\begin{equation*}
\bra{\Psi(A)}O\ket{\Psi(A)} = \diagramm{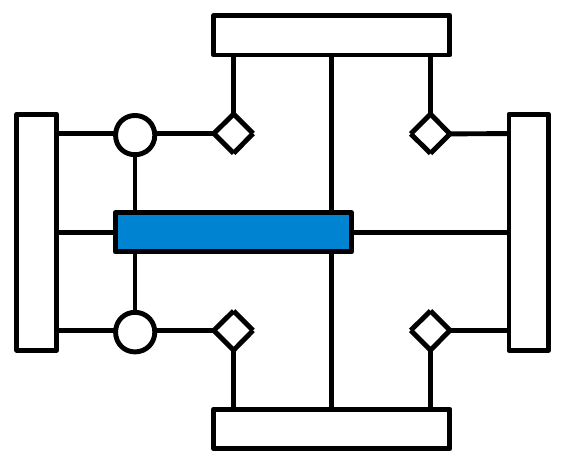},
\end{equation*} 
where the two-site operator can of course be oriented in the other channels as well.
\par The real power of the channel environment is now that arbitrary two-point correlation functions can be computed straightforwardly. Indeed, the expectation value of two operators at generic locations in the lattice is computed as
\begin{equation*}
\diagramm{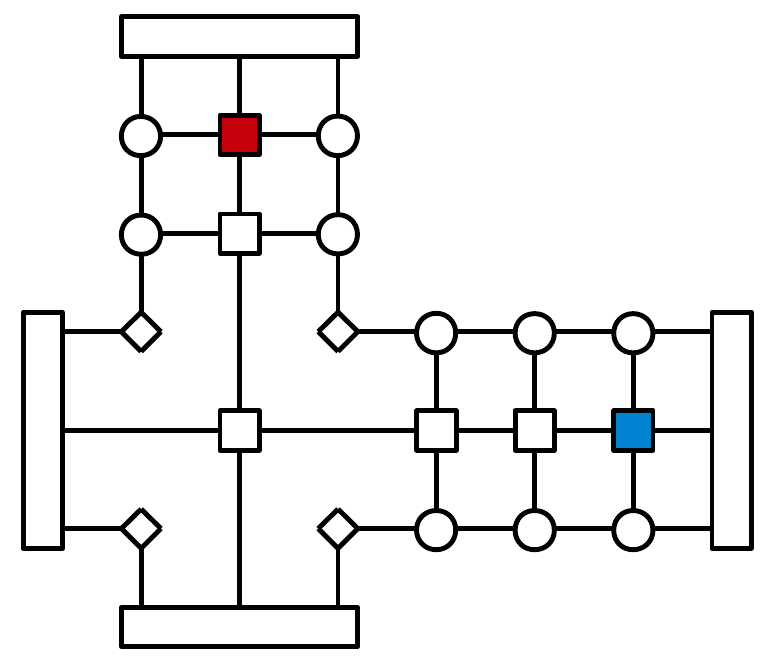}.
\end{equation*}
In fact, even three-point correlation functions can be evaluated by orienting the corners in the right way, as in, e.g., the contraction
\begin{equation*}
\diagramm{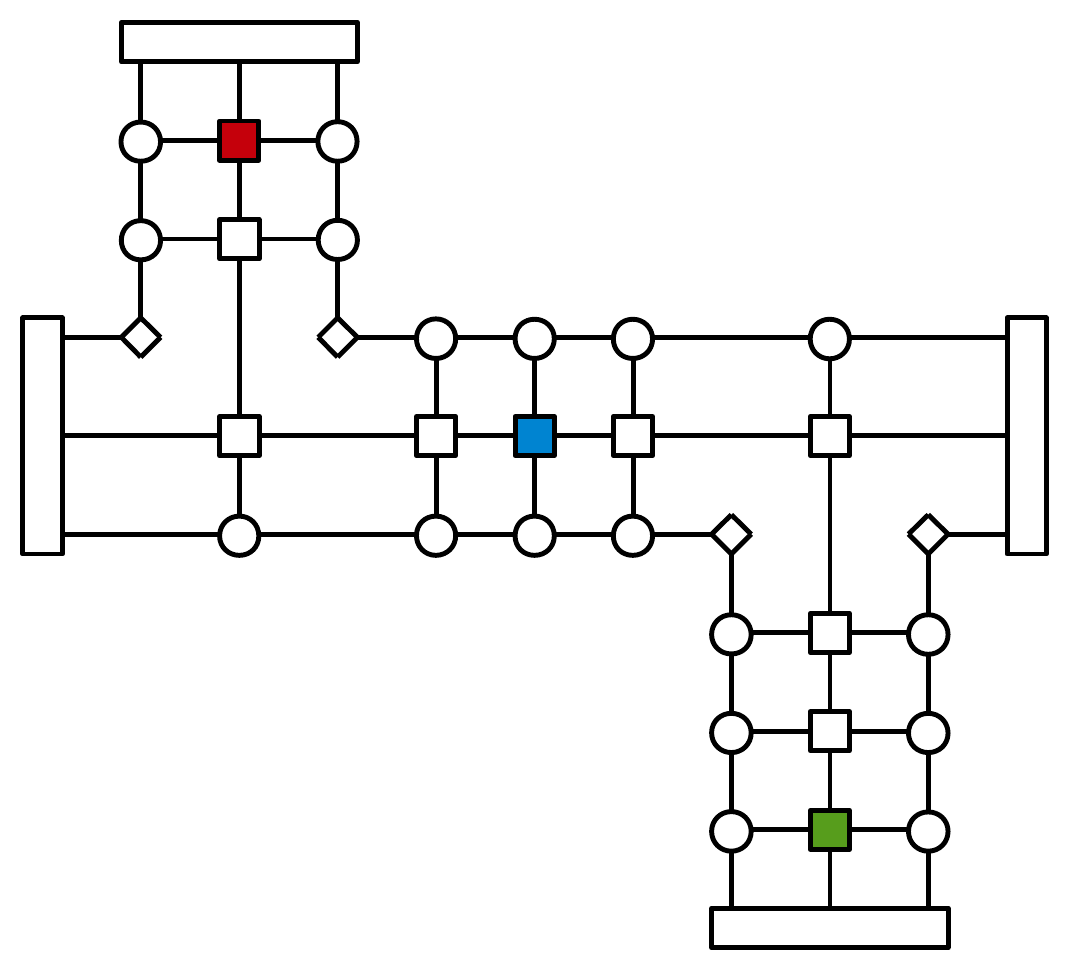}.
\end{equation*}
\par From a computational point of view, the hardest step in determining this corner environment is finding the fixed point of the linear transfer matrix; state-of-the-art algorithms \cite{prepJutho} scale as $\mathcal{O}(\chi^3D^4+\chi^2D^6)$, with $D$ the PEPS bond dimension and $\chi$ the bond dimension of the fixed point. In the case of strongly-correlated PEPS, finding the fixed point might take a lot of iterations. Determining the corner tensors and the channel fixed points has similar scalings, but this has to be done only once.

\subsection{Static structure factor}
\label{sec:sf}

As an example of the power of the corner environment, we will explicitly show how to compute a static correlation function directly in momentum space, i.e., the static structure factor $s(\vec{q})$,
\begin{equation*}
s(\vec{q}) = \frac{1}{|\mathcal{L}|} \sum_{i,j\in\mathcal{L}} \e^{i\vec{q}\cdot(\vec{n}_i-\vec{n}_j)} \bra{\Psi(A)} O_i\dag O_{j} \ket{\Psi(A)}_c
\end{equation*}
where only the connected part is taken up in the correlator, or, equivalently, the operators have been redefined such that their ground-state expectation value is zero. 
\par The momentum superposition of all relative positions of the operators can be evaluated explicitly by moving the operators independently through the channels and summing all contributions. This infinite number of contributions can be resummed by realizing that one obtains a geometric series inside the channels. Summing all different contributions from an operator moving in the top channel can be done by introducing a new momentum-resolved operator that captures the momentum superposition,
\begin{align}
\diagramm{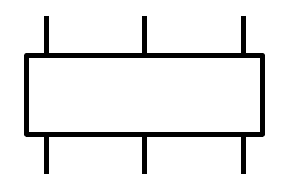} &= \sum_n \e^{iq_yn} \left( \diagramm{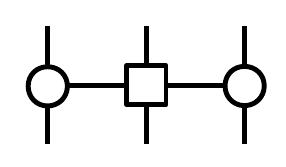}\right)^n \nonumber\\
&= \left[ 1 - \e^{iq_y} \left( \diagramm{structure5} - \diagramm{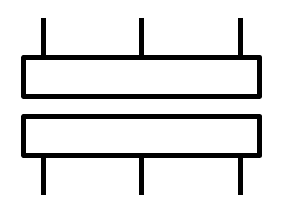} \right) \right]^{-1} \label{inverse}  \\
& \qquad + 2\pi\delta(q_y) \times \left( \diagramm{structure6} \right) \nonumber,
\end{align}
where we have separated the projector onto the largest eigenvector. As we will see, the diverging $\delta$ contribution will always drop out, such that the inverse is well defined. The momentum superposition inside the channel can be represented as
\begin{multline*}
\e^{iq_y} \diagramm{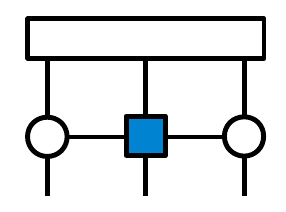} + \e^{2iq_y} \diagramm{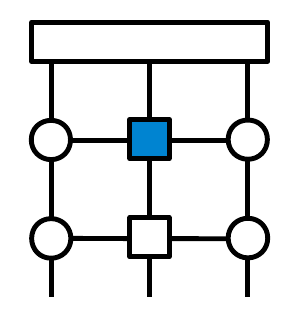} + \e^{3iq_y} \diagramm{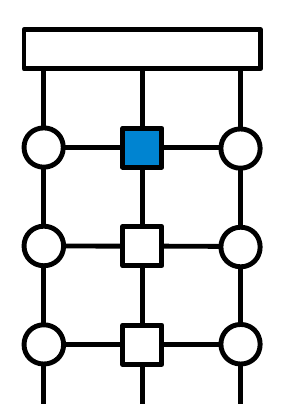} + \dots \\ = \e^{iq_y} \diagramm{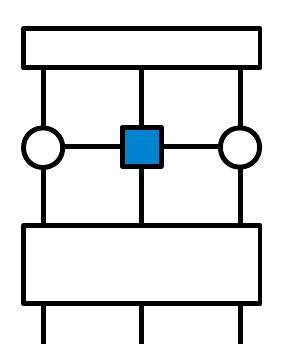},
\end{multline*}
where the component along the channel fixed point is indeed always zero---this component would correspond to the disconnected part of the correlation function. The geometric series converges for every value of the momentum and the inverse can be taken without problem.
\begin{widetext}
\par By independently letting the two operators travel through the channels all relative positions can be taken into account. In addition, we also need the contribution where the two operators act on the same site. The full expression is given by
\begin{align}
S(\vec{q}) &= \diagramm{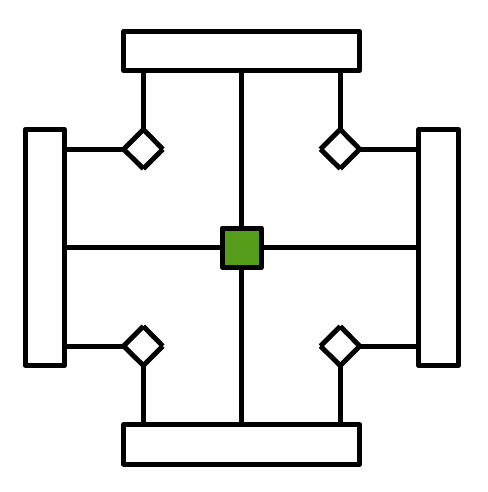} + \e^{-iq_x} \diagramm{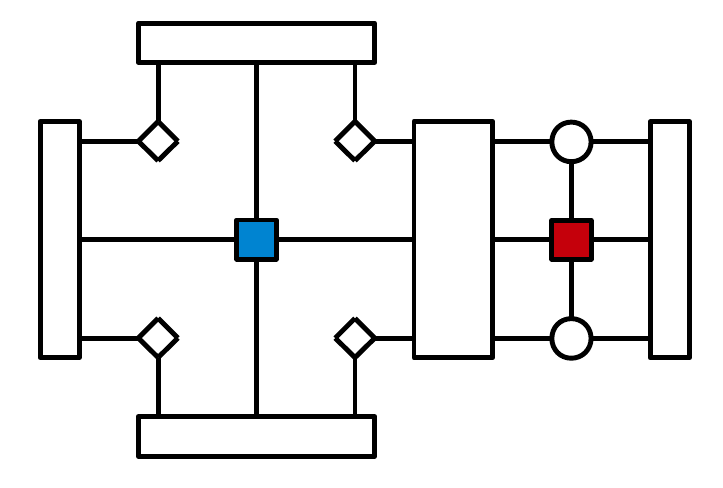} + \e^{+iq_x} \diagramm{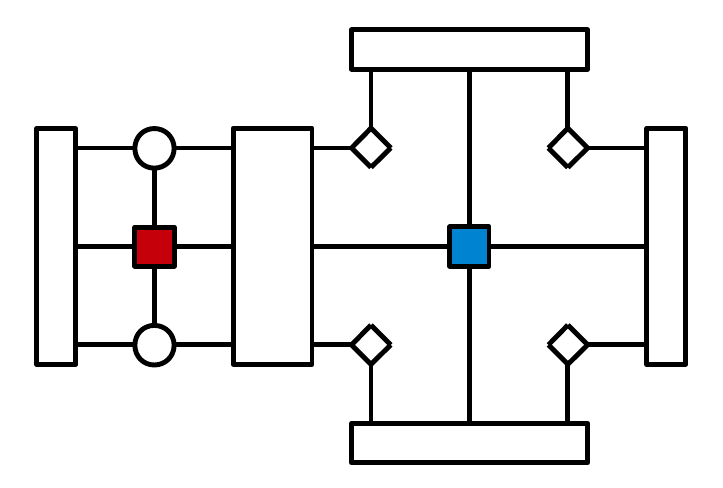} \nonumber \\
& \qquad + \e^{+iq_y} \diagramm{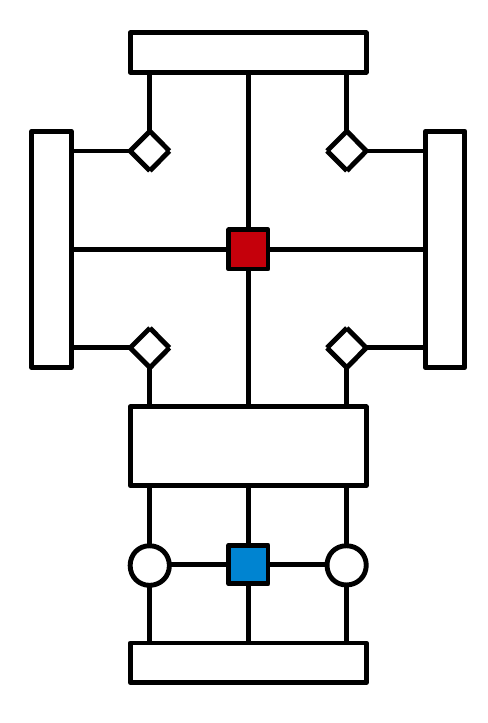} + \e^{-iq_y} \diagramm{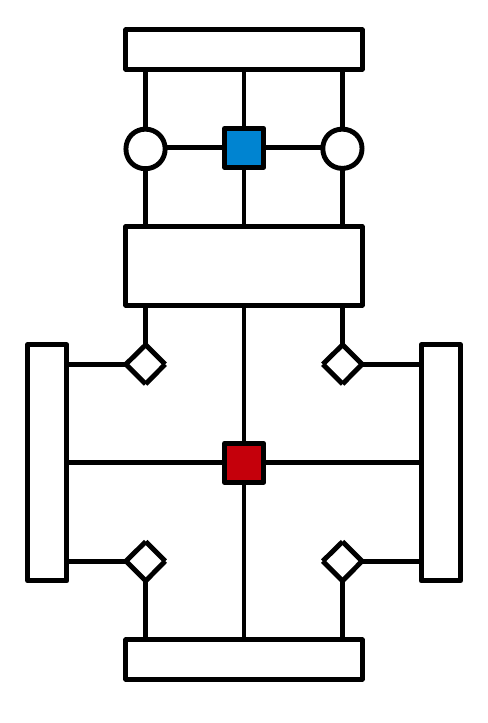} + \e^{+iq_x} \e^{-iq_y} \diagramm{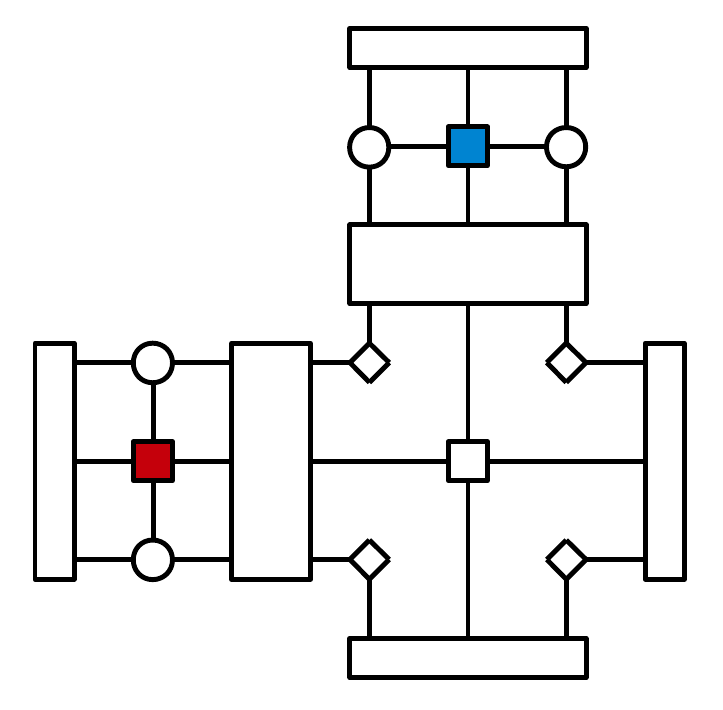} \nonumber \\
& \qquad + \e^{-iq_x} \e^{-iq_y} \diagramm{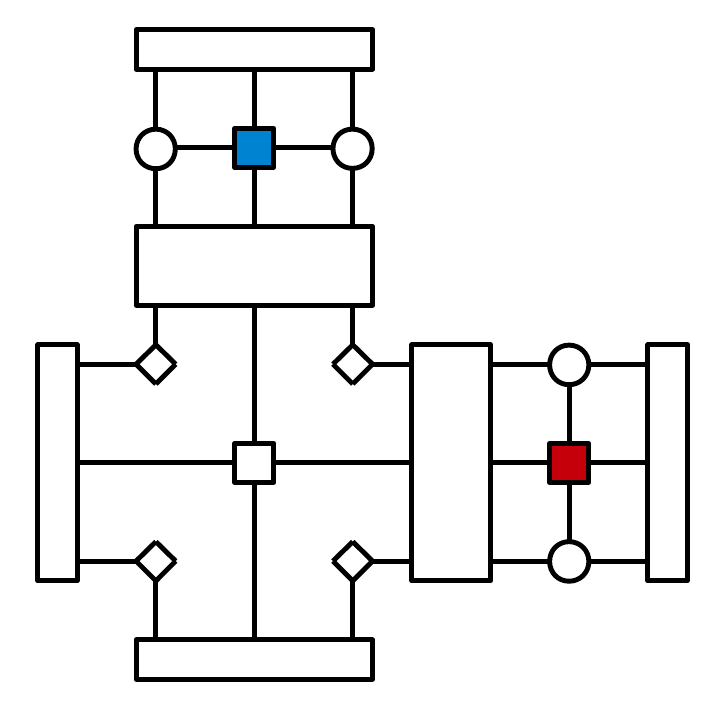}  + \e^{+iq_x} \e^{+iq_y} \diagramm{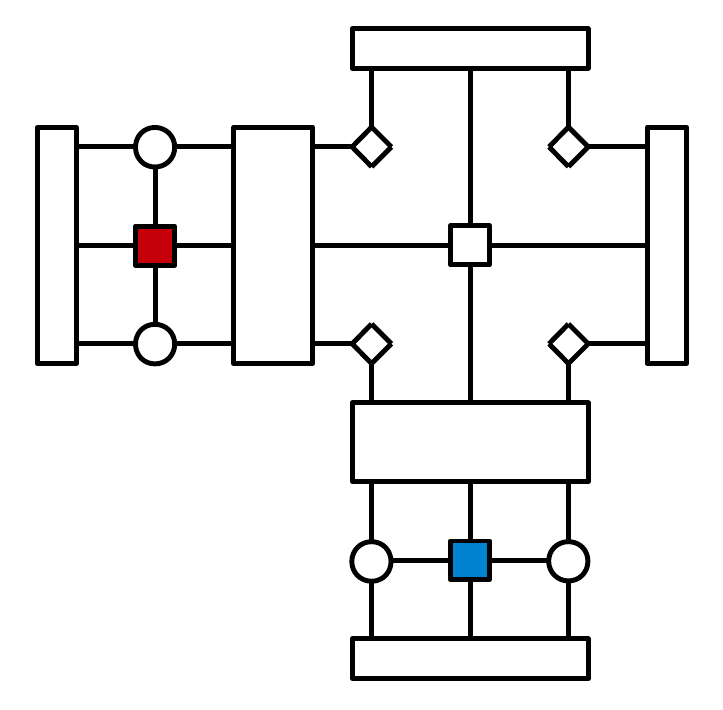} + \e^{-iq_x} \e^{+iq_y} \diagramm{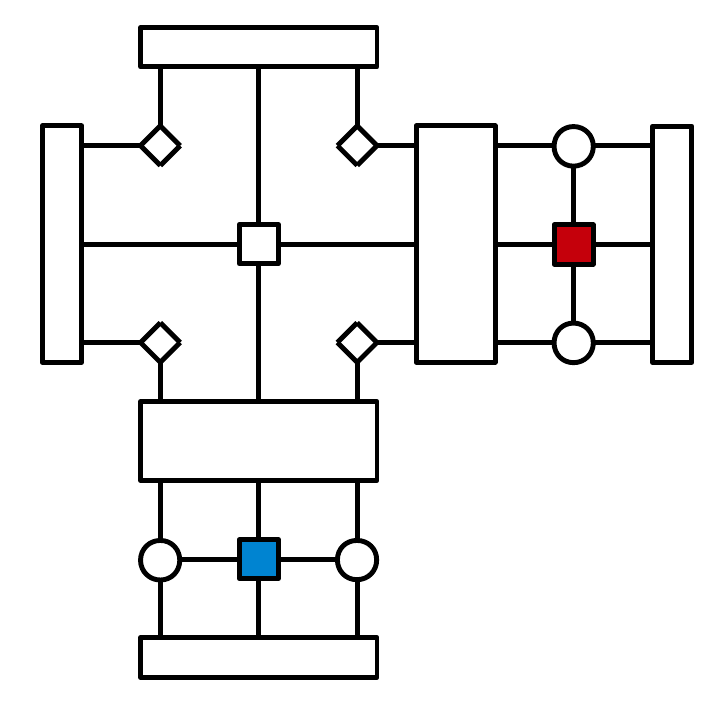} \label{sf},
\end{align}
where the green tensor represents the action of the two operators at the same site, and the blue and red tensors represent actions of the operators on the ket and bra level. The computational complexity for evaluating the structure factor scales as $\mathcal{O}(\chi^3D^4+\chi^2D^6)$ in the PEPS bond dimension $D$ and the bond dimension of the environment $\chi$, where the hardest step is computing the infinite sum inside a channel by an iterative linear solver.

\end{widetext}


\section{Variational conjugate-gradient method}
\label{sec:conj}

The PEPS ansatz defines a variational class of states that should approximate the ground state of two-dimensional quantum lattice systems in the thermodynamic limit. The system is described by its Hamiltonian, which we assume to consist of nearest-neighbor interactions, i.e.,
\begin{equation*}
H = \sum_{\braket{ij}} h_{ij},
\end{equation*}
and the lattice structure, for which we will confine ourselves to the square lattice. The following can be straightforwardly extended to different lattices, larger unit cells, or longer-range Hamiltonians.
\par As dictated by the variational principle, finding the best approximation to the ground state of $H$ now amounts to solving the highly non-linear minimization problem
\begin{equation} \label{optim}
\min_A \frac{\bra{\Psi(A)} H \ket{\Psi(A)}}{\braket{\Psi(A)|\Psi(A)}}.
\end{equation}
As we have seen, the evaluation of this energy functional for a certain tensor $A$ is already non-trivial, but can be done efficiently using a variety of numerical methods. Yet the evaluation of the energy is not enough, as efficient numerical optimization algorithms also rely on the evaluation of the gradient or higher-order derivatives of the energy functional. For a translation-invariant PEPS, the gradient is a highly non trivial object; it requires the evaluation of the change in energy from a variation in the tensor $A$, for which the effect of local \textit{and} non-local contributions should be added. In this respect, it is quite similar to a zero-momentum structure factor, and can be evaluated using the channel environment that we introduced in Sec.~\ref{sec:env}. In Sec.~\ref{sec:gradient} we run through the different diagrams for the gradient's explicit evaluation, and show that it can be computed efficiently.
\par With the gradient, the easiest algorithm is the steepest-descent method, where in each iteration one minimizes the energy in the direction of the gradient. One iteration $i$ corresponds to an update of the $A$ tensor as
\begin{equation*}
A_{i+1} \rightarrow A_i + \alpha \tilde{A}_i
\end{equation*}
with $\tilde{A}_i=-g_i$ ($g_i$ is the gradient at iteration $i$). The value of $\alpha>0$ is determined with a line-search algorithm; we have used a simple bisection algorithm with an Armijo condition on the step size \cite{Bonnans2006}. The performance can be greatly enhanced by implementing a non-linear conjugate-gradient method, where the search direction is a linear combination of the gradient and the direction of the previous iteration:
\begin{equation*}
\tilde{A}_{i} = -g_i + \beta_i \tilde{A}_{i-1} .
\end{equation*}
For each non linear optimization problem, the parameter $\beta_i$ can be chosen from a set of different prescriptions \cite{Bonnans2006, Nocedal2006, Hager2006}. Here we have exclusively used the Fletcher-Reeves scheme \cite{Fletcher1964}, according to which 
\begin{equation*}
\beta_{i} = \frac{\|g_{i}\|^2}{\|g_{i-1}\|^2}.
\end{equation*}
\par Crucially, these algorithms have a clear convergence criterion: when the norm of the gradient is sufficiently small, the energy cannot be further optimized and an optimal solution has been found.
\par Note that these direct optimization methods allow us to control the number of variational parameters in and/or impose certain symmetries on the PEPS tensor $A$: the iterative search can be easily confined to a certain subspace of the PEPS variational class by, e.g., projecting the gradient onto this subspace in each iteration. Moreover, this direct optimization strategy allows to start from a random input tensor $A$ and systematically converge to an optimal solution---all the results in Sec.~\ref{sec:bench} were obtained by starting from a random initial tensor.

\subsection{Computing the gradient}
\label{sec:gradient}

The objective function $f$ that we want to minimize [see Eq.~\ref{optim}] is a real function of the complex-valued $A$, or, equivalently, the independent variables $A$ and $\bar{A}$. The gradient is then obtained by differentiating $f(\bar{A},A)$ with respect to $\bar{A}$,
\begin{align*}
\text{grad} &= 2 \times \frac{\partial f(\bar{A},A) }{ \partial \bar{A} } \\
&= 2\times \frac{\partial_{\bar{A}} \bra{\Psi(\bar{A})}H\ket{\Psi(A)}  } {\braket{\Psi(\bar{A})|\Psi(A)}} \\
& \hspace{1cm} - 2\times \frac{\bra{\Psi(\bar{A})}H\ket{\Psi(A)}} {\braket{\Psi(\bar{A})|\Psi(A)}^2} \partial_{\bar{A}} \braket{\Psi(\bar{A})|\Psi(A)},
\end{align*}
where we have clearly indicated $A$ and $\bar{A}$ as independent variables. In the implementation we will always make sure the PEPS is properly normalized, such that the numerators drop out. By subtracting from every term in the Hamiltonian its expectation value, the full Hamiltonian can be redefined as
\begin{equation} \label{subtract}
H \rightarrow H - \bra{\Psi(\bar{A})} H \ket{\Psi(A)},
\end{equation}
such that the gradient takes on the simple form
\begin{align*}
\text{grad} &= 2 \times \partial_{\bar{A}} \bra{\Psi(\bar{A})}H\ket{\Psi(A)} .
\end{align*}
The gradient is thus obtained by differentiating the energy expectation value $\bra{\Psi(\bar{A})}H\ket{\Psi(A)}$ with respect to every $\bar{A}$ tensor in the bra level and taking the sum of all contributions. Every term in this infinite sum is obtained by omitting one $\bar{A}$ tensor and leaving the indices open. The full infinite summation is then obtained by letting the Hamiltonian operator and this open spot in the network travel through the channels separately, just as in the case of the structure factor in Sec.~\ref{sec:sf}.
\par Let us first define a new tensor that captures the infinite sum of Hamiltonian operators acting inside a channel,
\begin{align*}
\diagramm{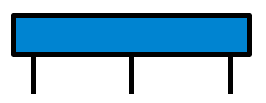} &= \diagramm{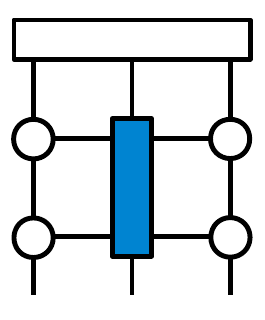} + \diagramm{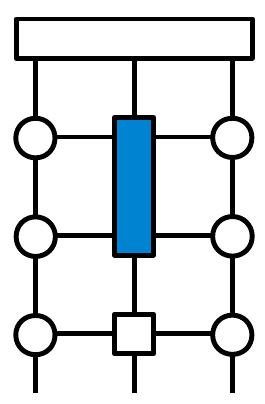} + \diagramm{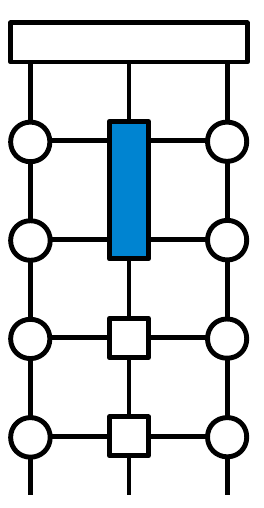} + \dots \\
&= \diagramm{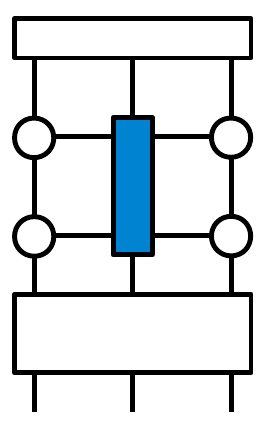},
\end{align*}
where the big tensor is again the inverted channel operator of Eq.\eqref{inverse} with momentum zero. Because we have redefined the Hamiltonian in Eq.~\eqref{subtract}, the inversion of the channel operator is well defined, because the vector on which the inverse acts has a zero component along the channel fixed point $\rho_L$. 
\par With this blue tensor all different relative positions of the Hamiltonian terms and the tensor $\bar{A}$ that is being differentiated (the open spot) can be explicitly summed, similarly to the expression for the structure factor [Eq.~\ref{sf}]. There are a few more terms because every Hamiltonian term corresponds to a two-site operator and has different orientations.
\begin{widetext}
\par The full expression is
\begin{align*} 
\text{grad} &= \diagramm{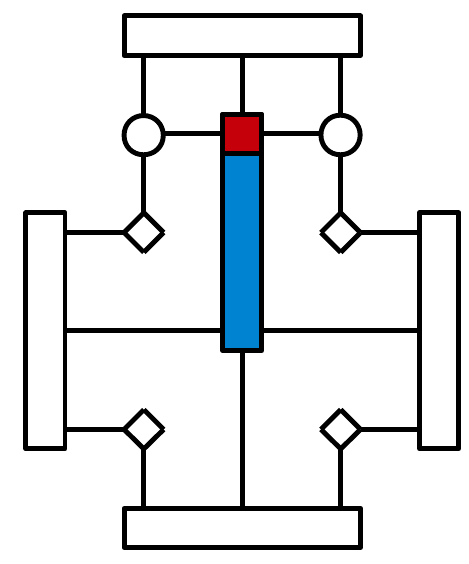} + \diagramm{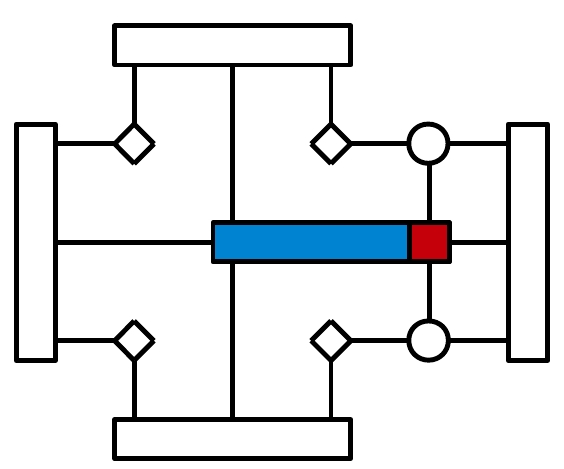} + \diagramm{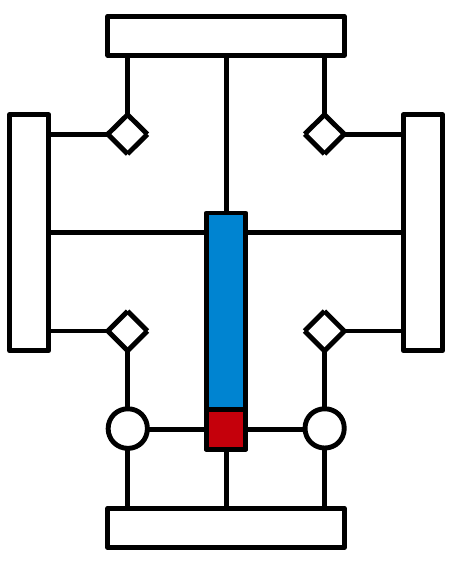} + \diagramm{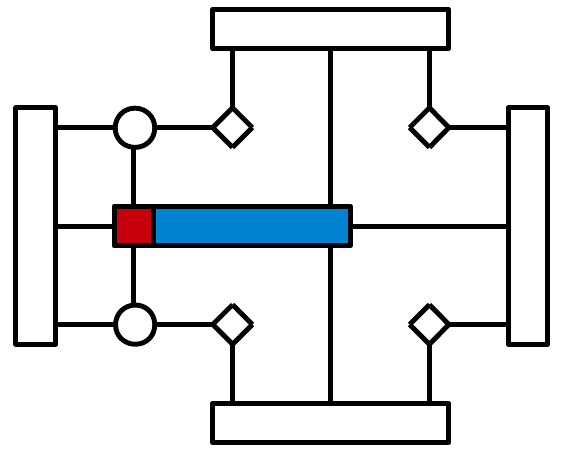} \\
& \qquad + \ \diagramm{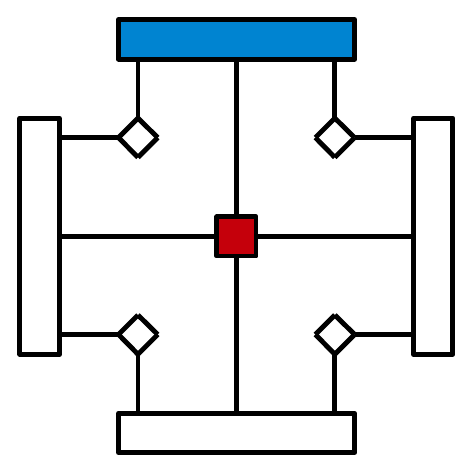} + \diagramm{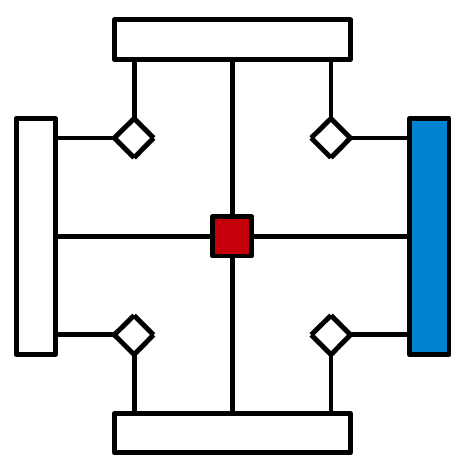} + \diagramm{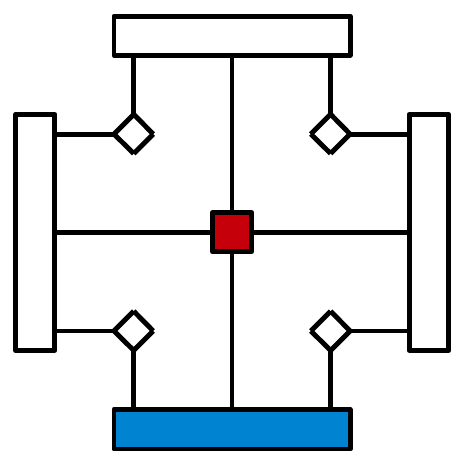} + \diagramm{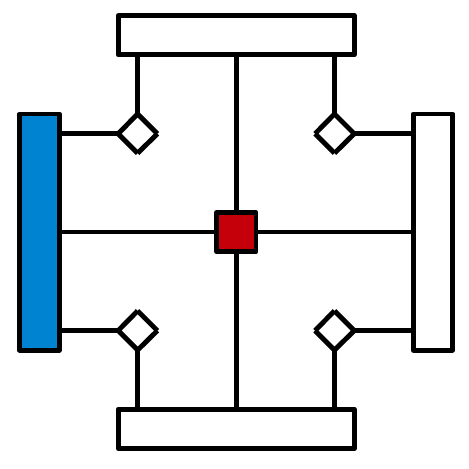} \\
& \qquad + \diagramm{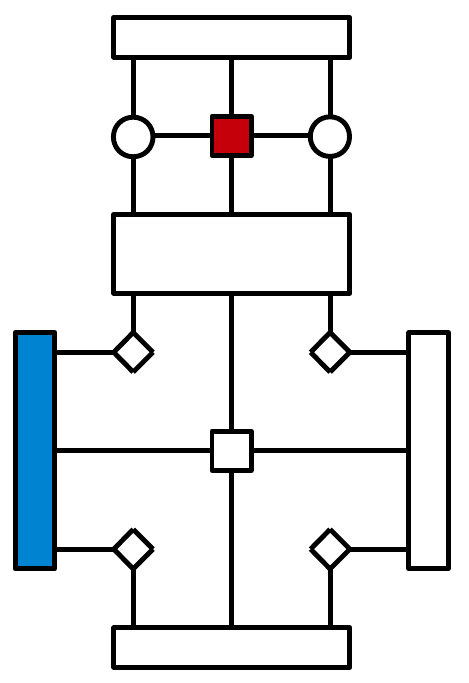} + \diagramm{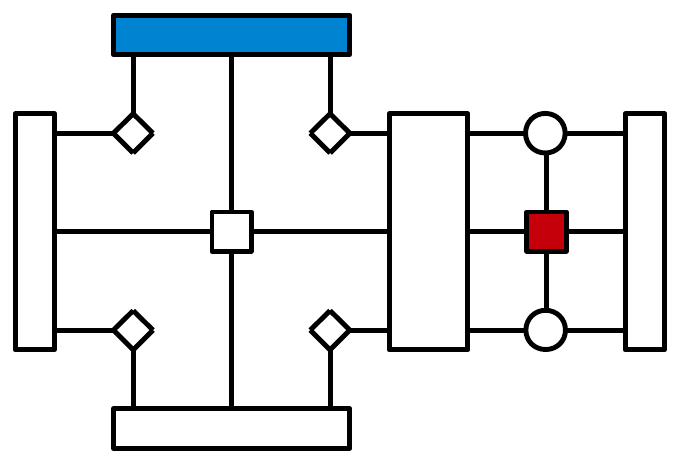} + \diagramm{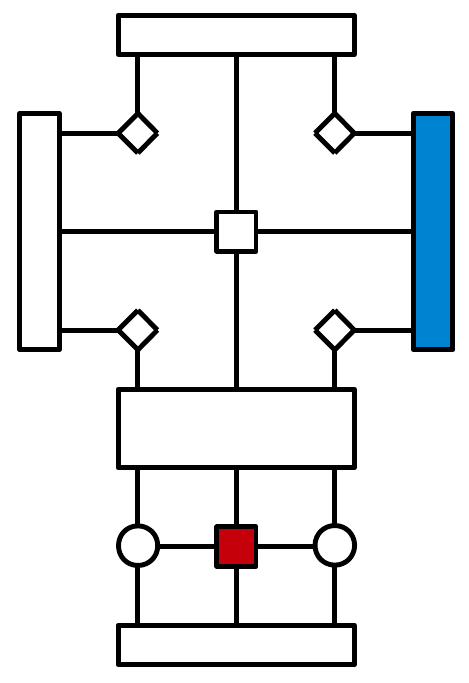} + \diagramm{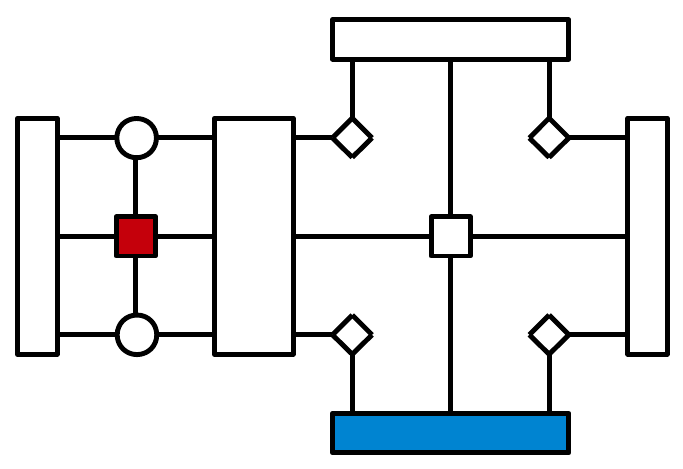} \\
& \qquad + \diagramm{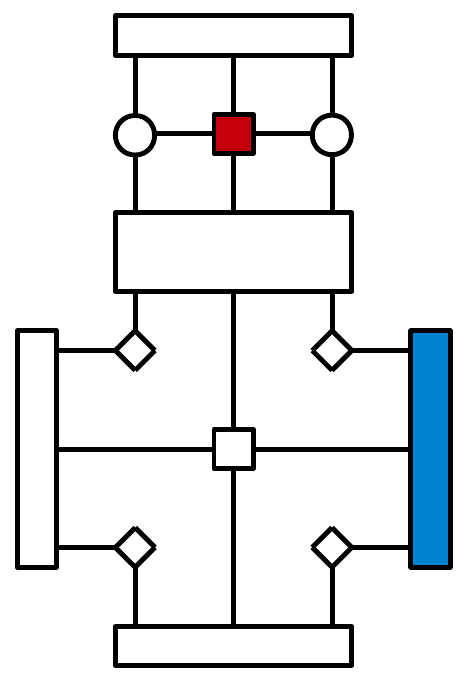} + \diagramm{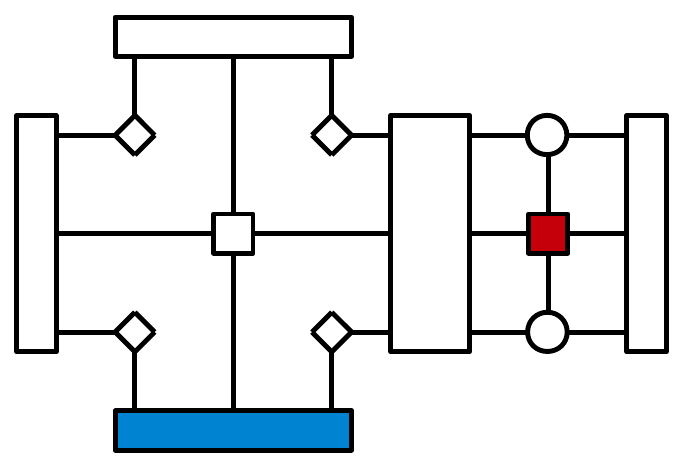} + \diagramm{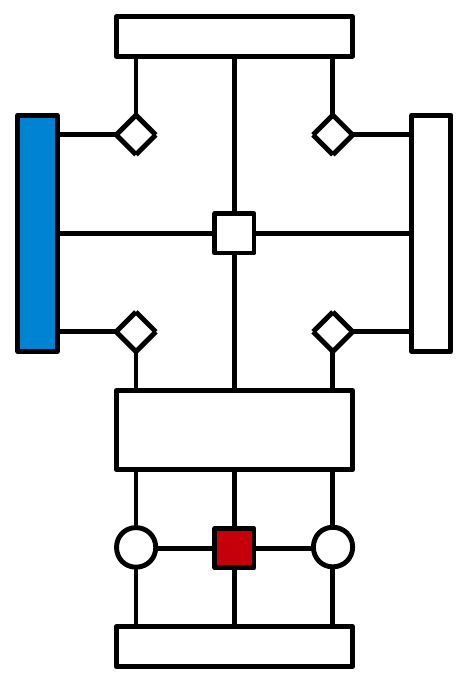} + \diagramm{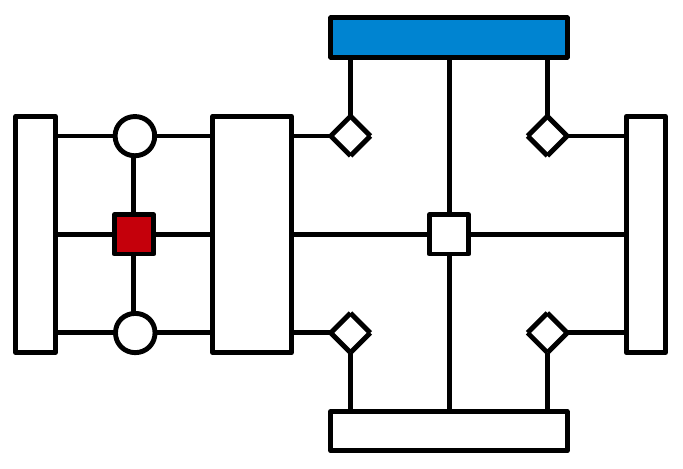} \\
& \qquad + \diagramm{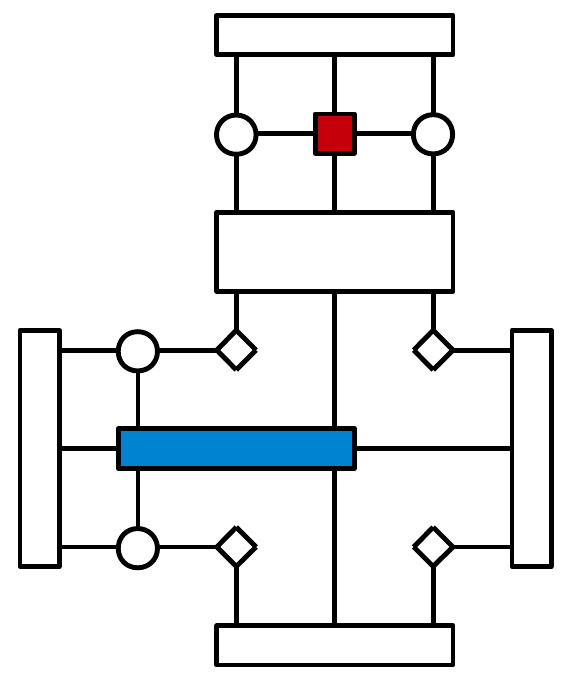} + \diagramm{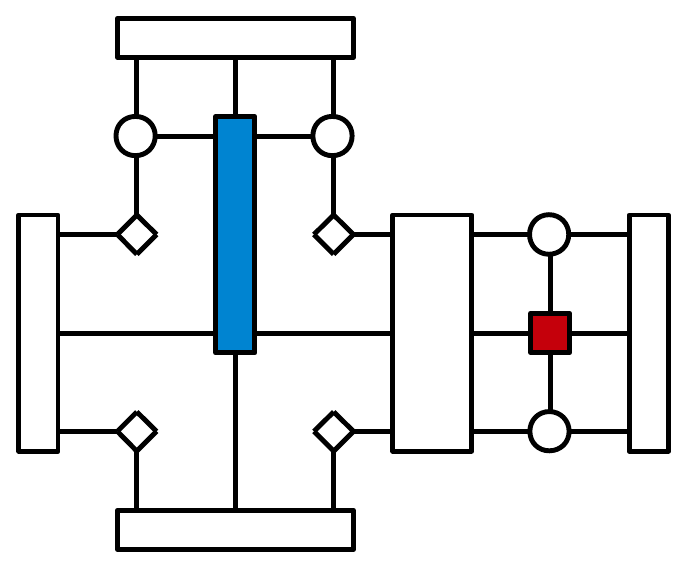} + \diagramm{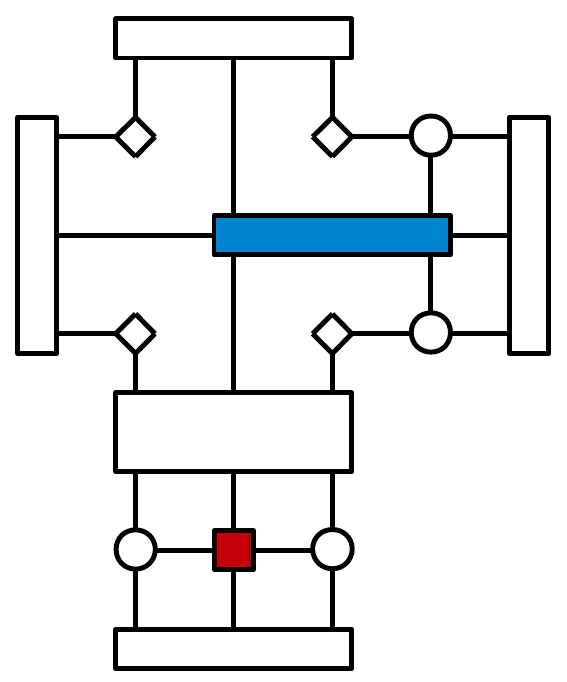} + \diagramm{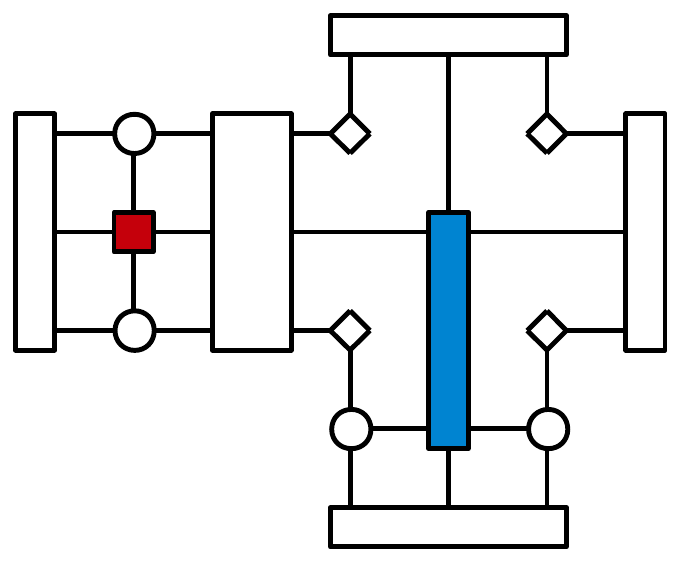} \\
& \qquad + \diagramm{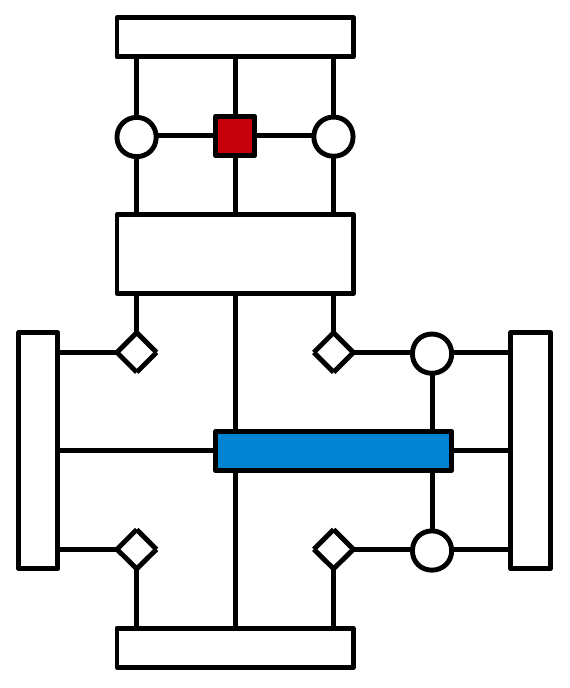} + \diagramm{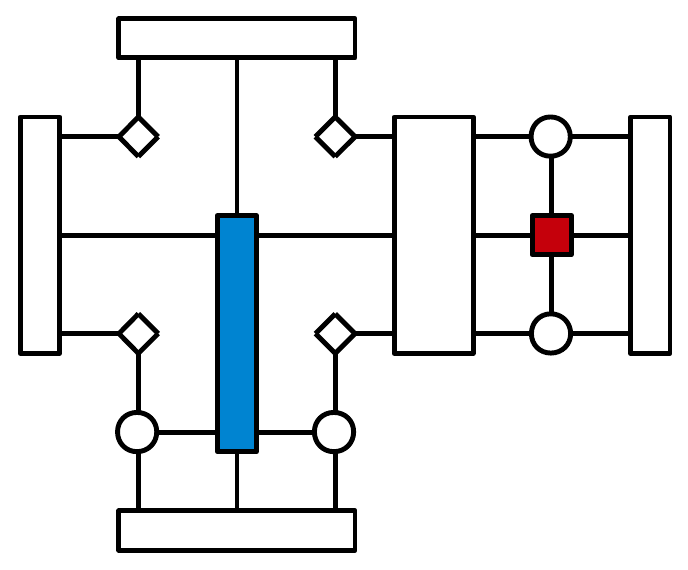} + \diagramm{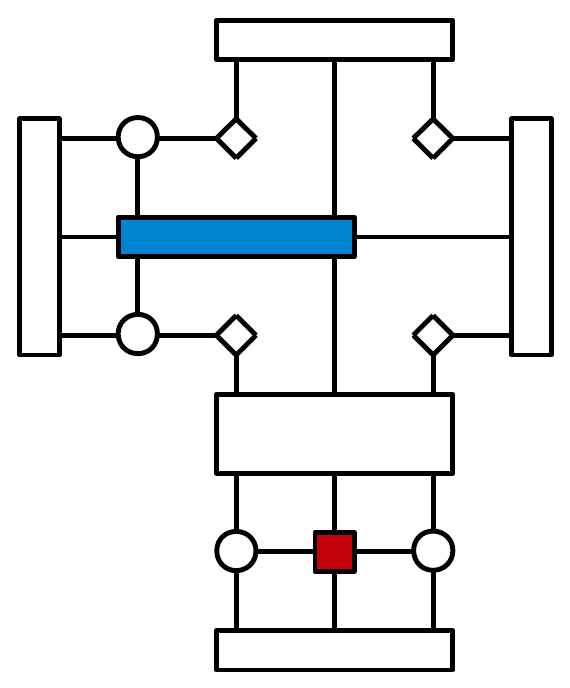} + \diagramm{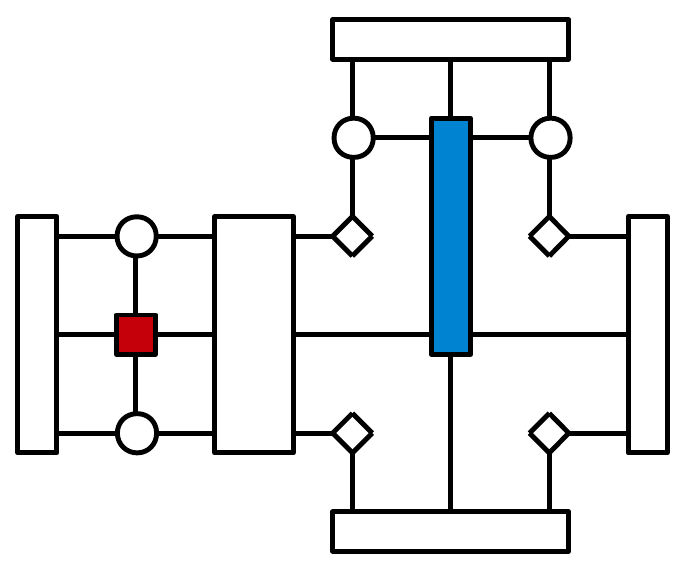},
\end{align*}
where the red tensor indicates where the open spot in the bra level of the diagram is. Note that the diagrams on the same line are always related by a rotation; in the case that the PEPS tensor $A$ is rotationally invariant, these diagrams give exactly the same contribution. This implies that the gradient corresponding to a rotationally invariant tensor $A$ is itself rotationally invariant. The computational complexity for evaluating the gradient scales similarly to the structure factor, i.e. $\mathcal{O}(\chi^3D^4+\chi^2D^6)$, where again the hardest step is computing the infinite sum inside a channel by an iterative linear solver.
\end{widetext}

\subsection{The energy variance}
\label{sec:variance}

Like any variational method, the PEPS ansatz is a priori not guaranteed to provide an accurate parametrization of a ground state. It is expected that increasing the PEPS bond dimension provides a good test for the reliability of the simulation: an extrapolation in $D$ should provide the correct results. One problem is that it is unclear how the energy or order parameter behave as a function of $D$ \cite{Corboz2016}. A better and completely unbiased extrapolation quantity is the energy variance \cite{Sorella2001}, defined as
\begin{equation*}
v = \bra{\Psi(A)} \left( H-e \right)^2 \ket{\Psi(A)},
\end{equation*}
with $e=\bra{\Psi(A)}H\ket{\Psi(A)}$ the energy expectation value. It measures to what extent a variational wave function approximates the ground state (or more generally, an eigenstate) of the Hamiltonian. 
\par Because the variance can be interpreted as a zero-momentum structure factor of the Hamiltonian operator, the computation of the energy variance is again similar. In addition to the green tensor above, we will also need the following geometric series
\begin{align*}
\diagramm{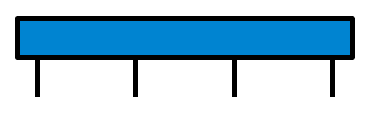} &= \diagramm{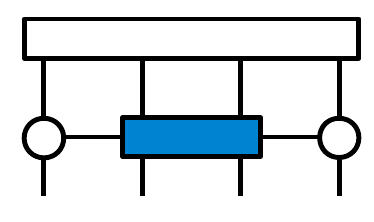} + \diagramm{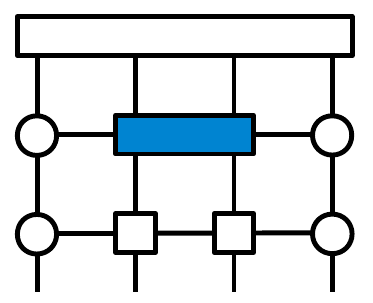} + \dots \\ 
&=  \diagramm{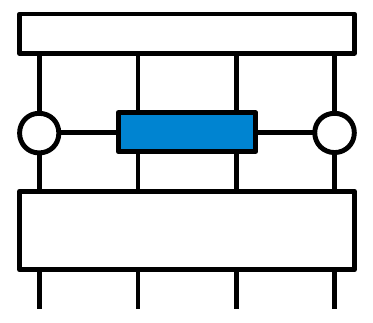}
\end{align*}
where
\begin{align*}
\diagramm{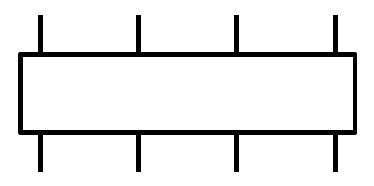} &=  \sum_n \left( \diagramm{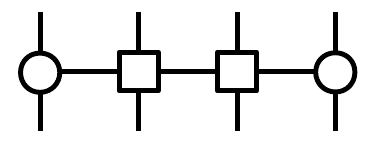}\right)^n \\
& = \left[ 1 - \left( \diagramm{var5} - \diagramm{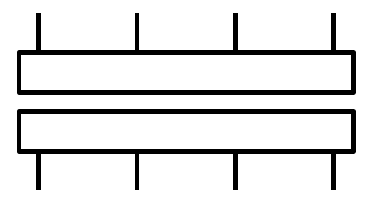} \right) \right]^{-1} \\
& \qquad + 2\pi\delta(0) \times \left(\diagramm{var6}\right)
\end{align*}
with the fixed points of the two-site channels properly normalized. We again renormalize the Hamiltonian as
\begin{equation*}
H \rightarrow H - \bra{\Psi(A)} H \ket{\Psi(A)}
\end{equation*}
such that disconnected contributions always drop out and the inverse of the operator above is well defined. The blue tensor has $\chi^2D^4$ elements, so its computation is by far the most costly step for the variance evaluation. Approximating it by a tensor decomposition might reduce the cost considerably, but for our purposes this has not been necessary.
\par Let us now associate to each nearest-neighbor term $\braket{ij}$ in the Hamiltonian a variance term as
\begin{align*}
v_{\braket{ij}} = \bra{\Psi(A)} H h_{\braket{ij}} \ket{\Psi(A)},
\end{align*}
such that the energy variance per site is given by
\begin{equation*}
v = \frac{1}{|\mathcal{L}|} \bra{\Psi(A)} H^2 \ket{\Psi(A)} = v_{\braket{ij},\text{hor}} + v_{\braket{ij},\text{ver}},
\end{equation*}
the sum of the variances corresponding to the horizontal and vertical nearest-neighbour terms in the Hamiltonian.
\begin{widetext}
\par The vertical contribution is given by
\begin{align*}
v_{\braket{ij},\text{ver}} &= \diagramm{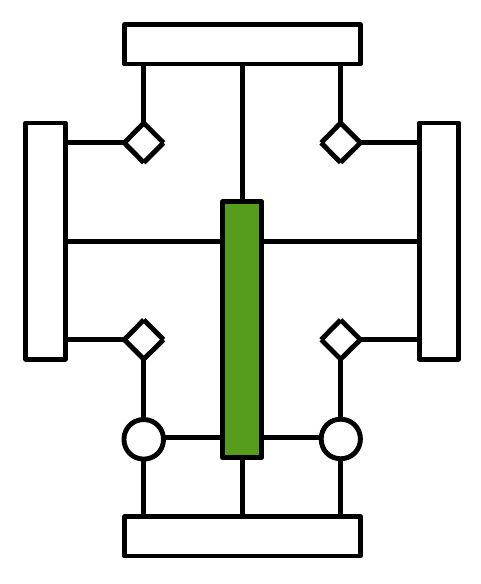} + 2 \times \diagramm{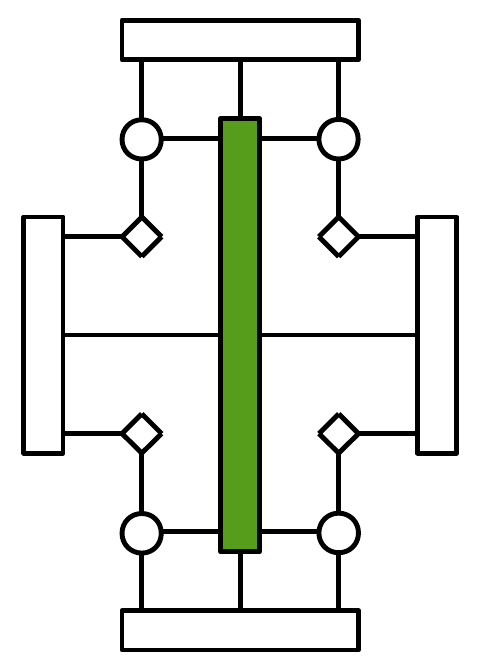} + 2 \times \diagramm{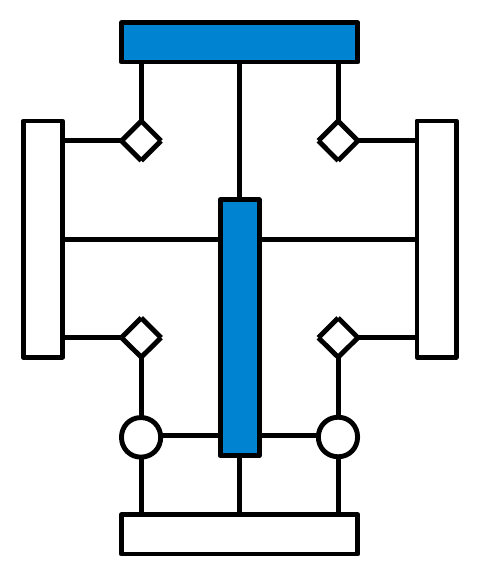} + 2 \times \diagramm{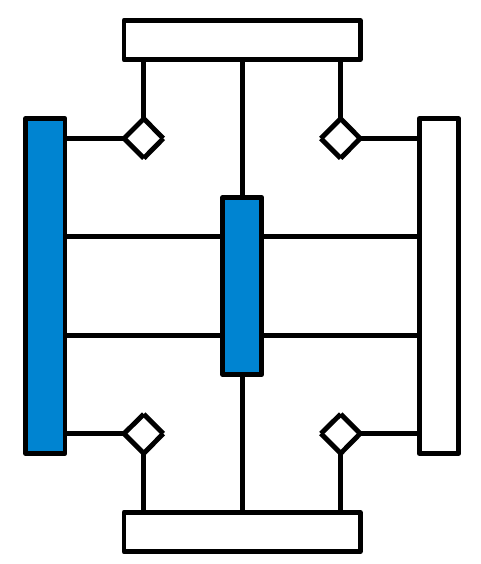} \\
& \qquad + 2\times \diagramm{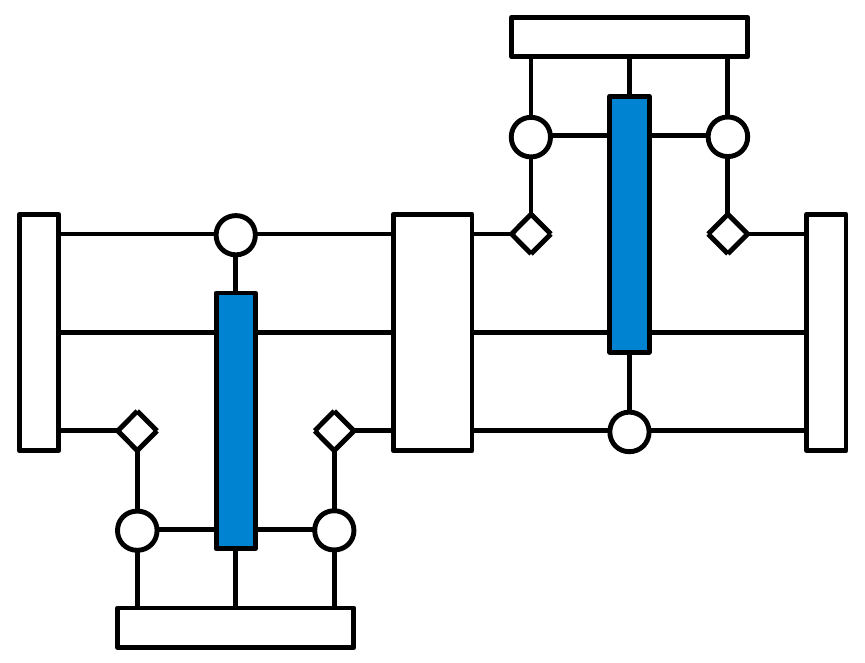} + 2\times \diagramm{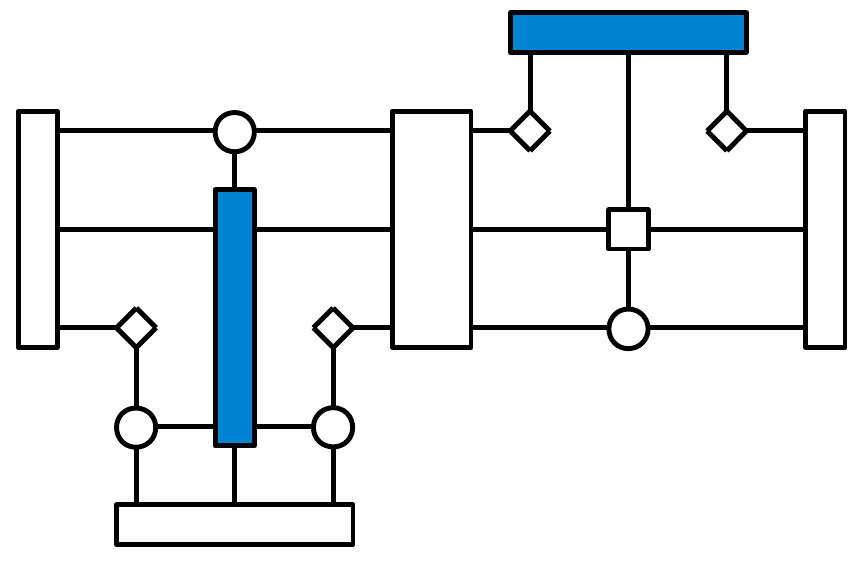} \\
& \qquad + 2\times \diagramm{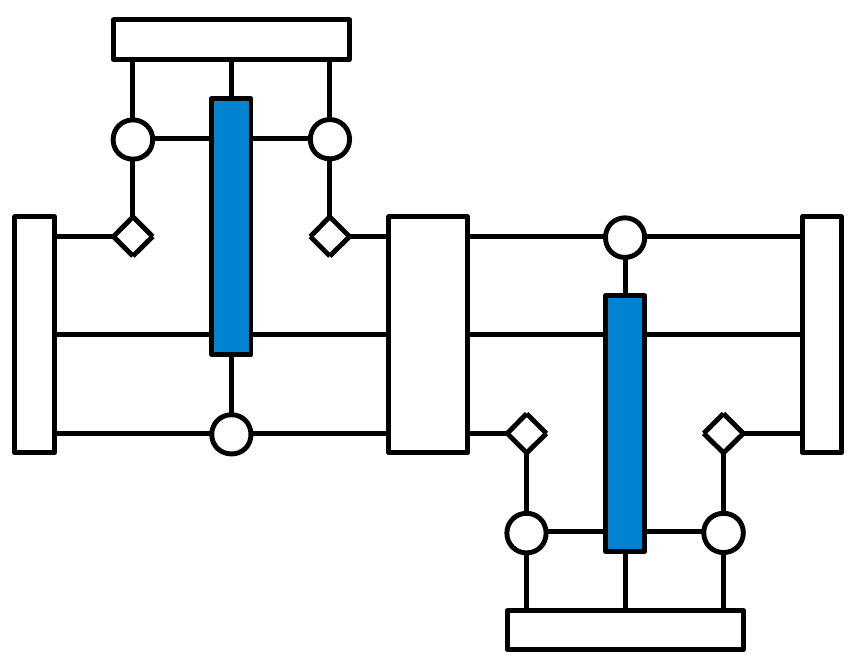} + 2\times \diagramm{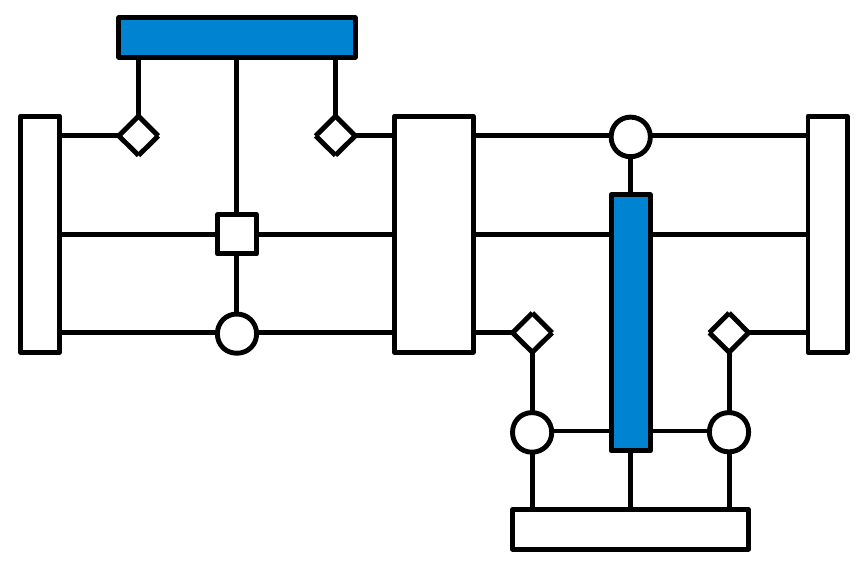}  \\
& \qquad + 2\times \diagramm{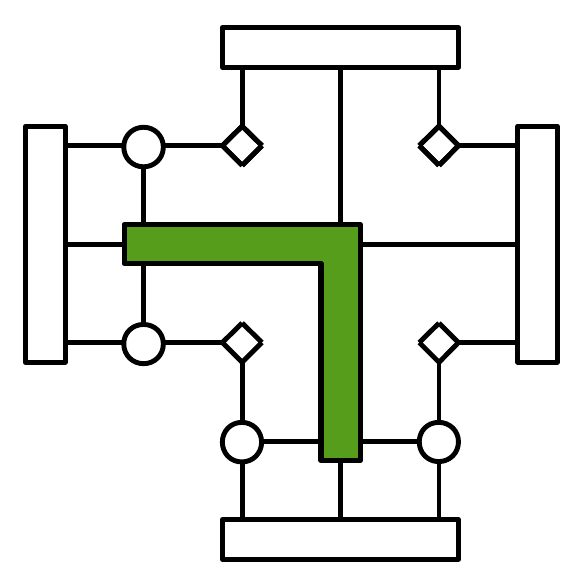} + 2\times \diagramm{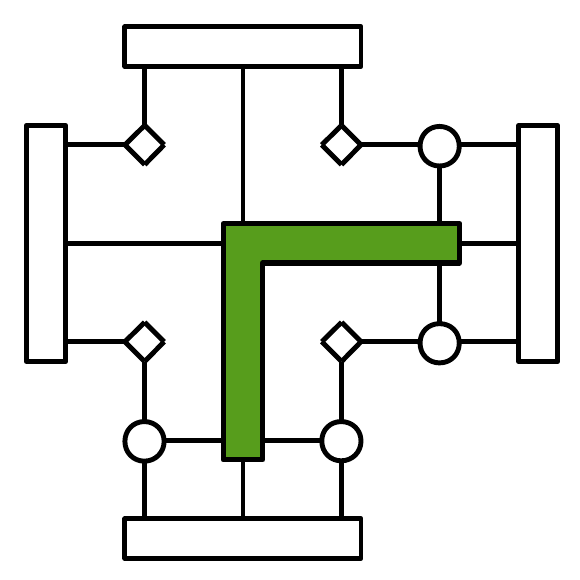} + 2\times \diagramm{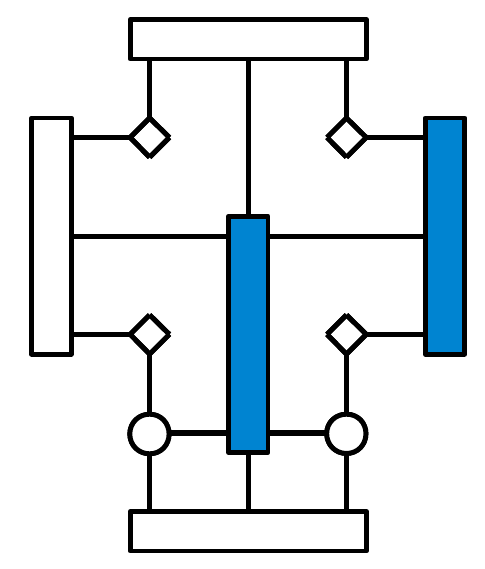} + 2\times \diagramm{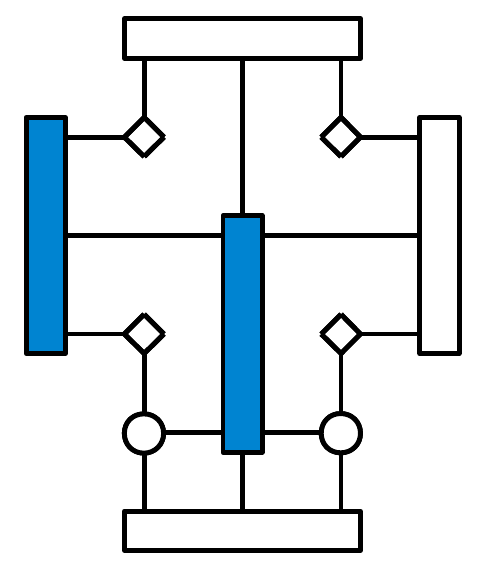} \\
& \qquad + 2\times \diagramm{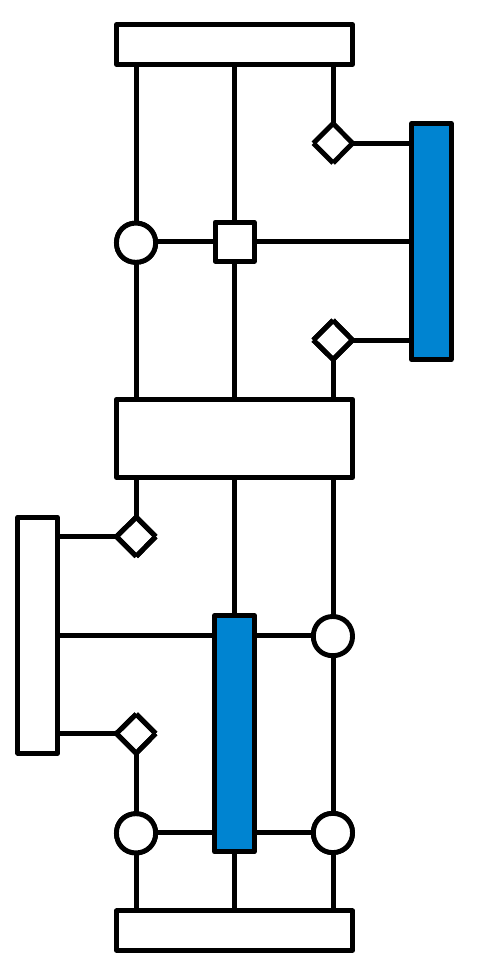} + 2\times \diagramm{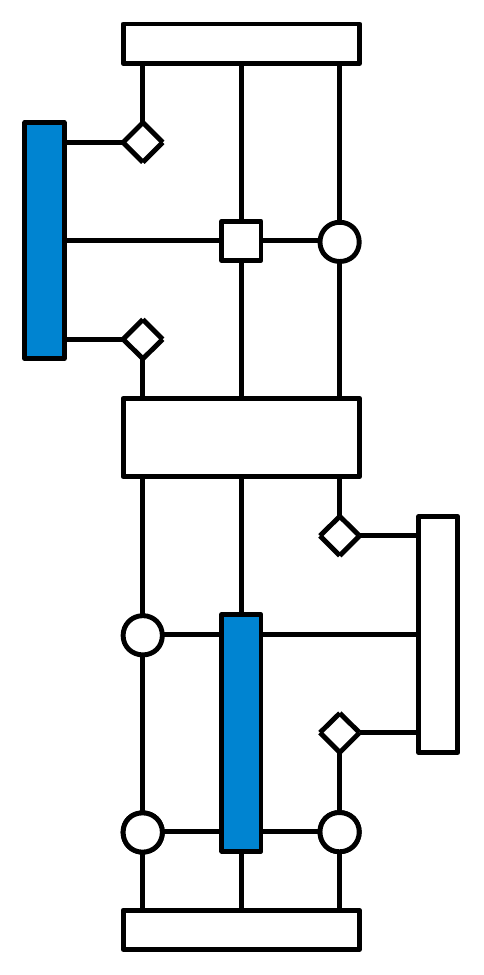}  + 2\times \diagramm{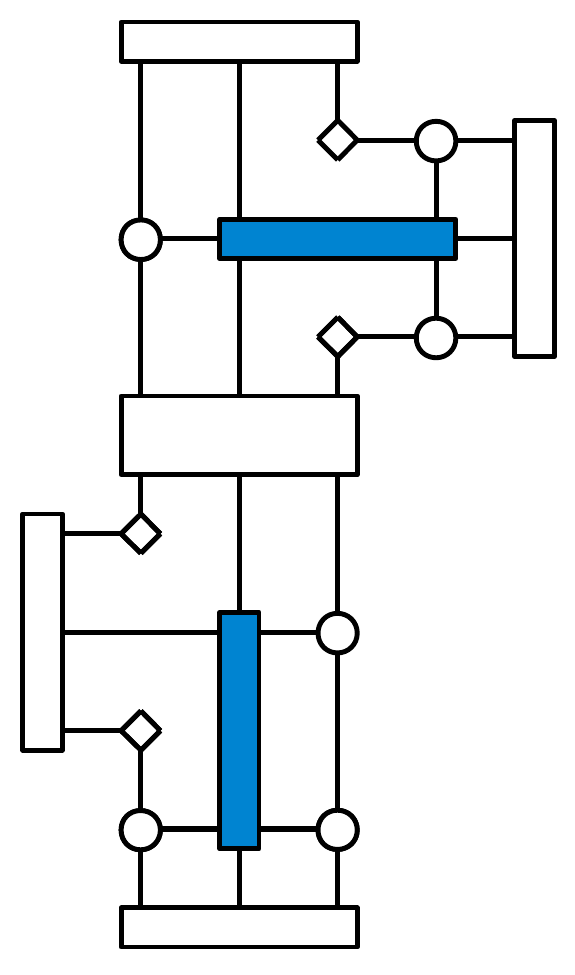} + 2\times \diagramm{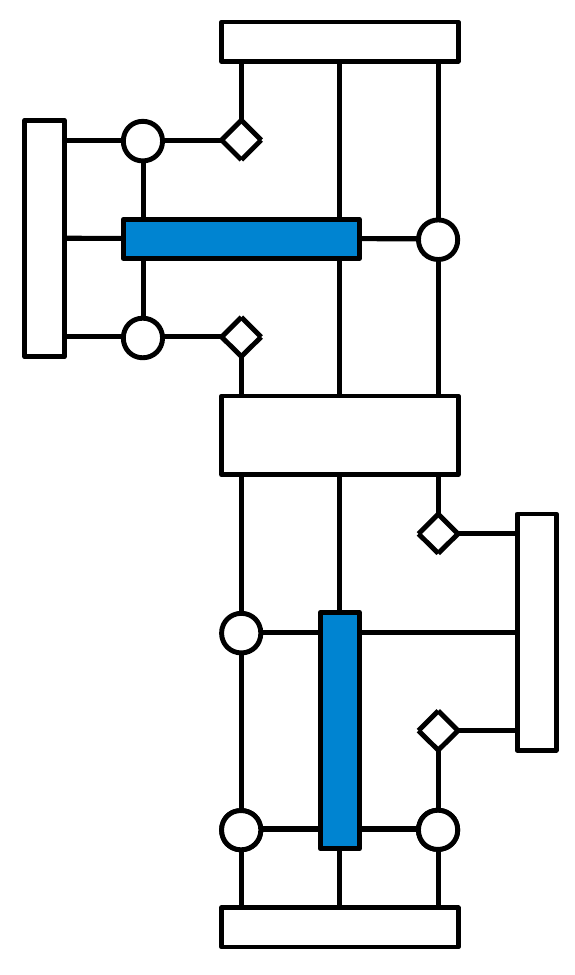} .
\end{align*}
The green tensors represent the double action of the Hamiltonian operator: a two-site tensor if they fully overlap and a three-site tensor if the overlap is on one site only. In this expression, we have explicitly used the rotational invariance of the PEPS tensor $A$, which can be easily imposed within our framework. Under this symmetry, the horizontal and vertical contributions to the variance are obviously equal, so the above is the complete expression for the variance. If $A$ is not rotationally invariant, all the other diagrams can be obtained by rotating the above ones. The complexity scaling of the variance evaluation is larger than for the gradient, because of the extra geometric series in a two-site channel; the complexity scales as $\mathcal{O}(\chi^3D^6)$.

\end{widetext}


\section{Benchmarks}
\label{sec:bench}

\par As a first check, we apply our PEPS algorithm to the two-dimensional transverse Ising model on the square lattice, defined by the Hamiltonian
\begin{equation*}
H_{\text{Ising}} = \sum_{\braket{ij}} S^z_i S^z_j + \lambda \sum_i S^x_i.
\end{equation*}
The model exhibits a phase transition at $\lambda_c\approx3.044$ \cite{Blote2002} from a symmetry broken phase to a polarized phase; the order parameter is $m=\braket{S^z}$. The model has been extensively studied with the PEPS ansatz \cite{Jordan2008, Orus2009, Phien2015}, and we use the model as a benchmark for our conjugate-gradient method. 
\par
\begin{figure}
\includegraphics[width=\columnwidth]{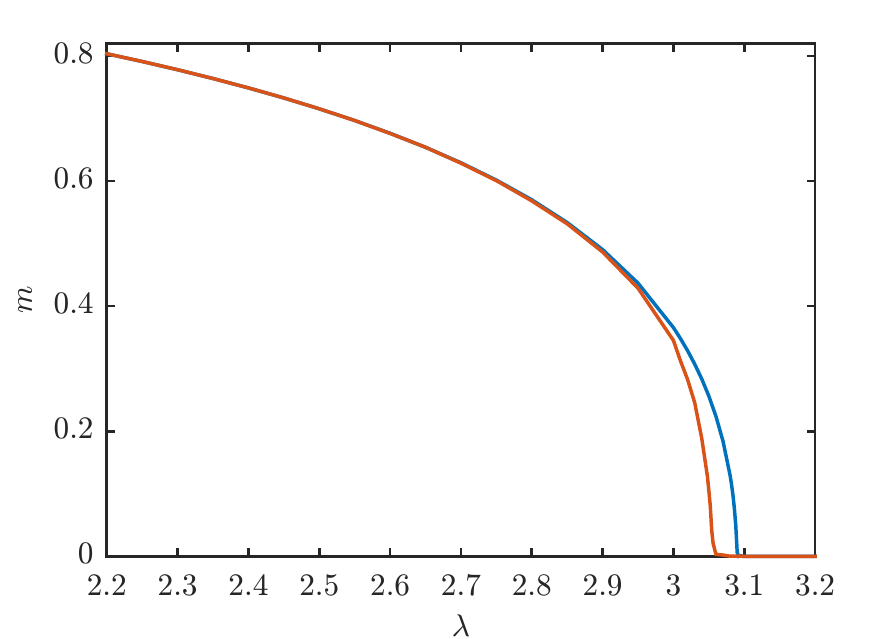}
\caption{The magnetization curve for the transverse Ising model with bond dimensions $D=2$ (blue) and $D=3$ (orange). We nicely capture the phase transition, although the critical point has been slightly shifted. The critical point can be estimated as the point where the slope of the curve is maximal; we arrive at $\lambda_c\approx3.09$ ($D=2$) and $\lambda_c\approx3.054$ ($D=3$).}
\label{fig:ising}
\end{figure}

 In Fig.~\ref{fig:ising} we have plotted the magnetization curve of the transverse Ising model for two different values of the bond dimension $D$, the parameter that controls the dimensions of the PEPS and can be tuned as a refinement parameter. We see that the phase transition is captured accurately already for $D=3$; growing the bond dimension further will increase the accuracy only slightly. Further on, we will observe that a systematic growing of the bond dimension is paramount for capturing  ground states with stronger correlations.
\par
\begin{figure}
\subfigure{\includegraphics[width=0.49\columnwidth]{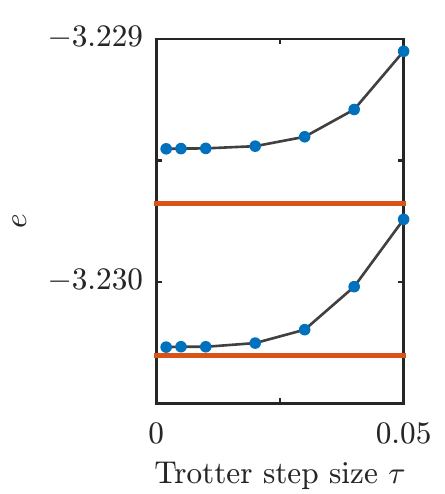}}
\subfigure{\includegraphics[width=0.49\columnwidth]{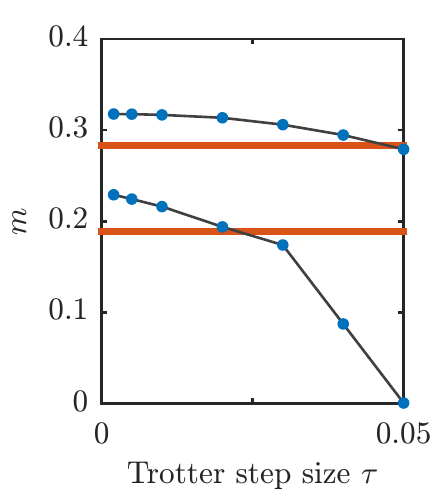}}
\caption{Our variational results compared to the results that are obtained with imaginary-time evolution using the full-update algorithm; the comparison is done for the transverse Ising model at $\lambda=3.04$ for bond dimensions $D=2$ and $D=3$. On the left we have plotted the convergence for the energy (magnetization) as a function of the Trotter step size of the full-update scheme (blue points), and our results (red line). For both plots, the upper (lower) lines are for $D=2$ ($D=3$).}
\label{fig:ite}
\end{figure}

 In Fig.~\ref{fig:ite} we have compared our variational search with imaginary-time evolution (full update), showing that we find lower energies and better order parameters, even as the Trotter error goes to zero. The plot clearly shows that, as the Trotter step size goes to zero, the imaginary-time result does not converge to the variational optimum that we obtain. Note that the variational freedom is slightly different: we optimize over a rotationally symmetric PEPS with a one-site unit cell, whereas the imaginary-time results break rotational symmetry and work with a two-site unit cell. Although this larger rotationally asymmetric unit-cell might give lower energies, it appears that our optimization still gives better energies and order parameters.
\par
\begin{figure}
\subfigure{\includegraphics[width=\columnwidth]{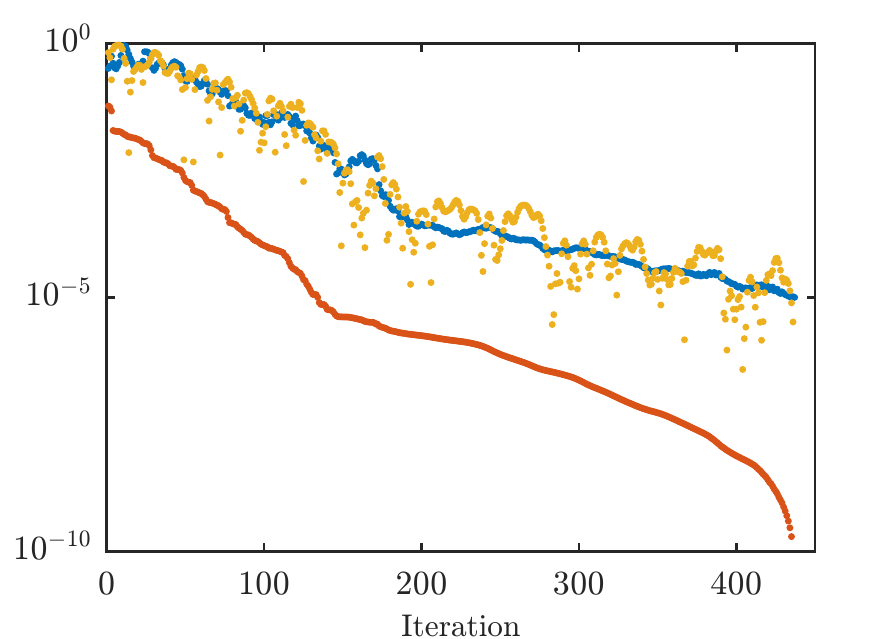}} \\
\subfigure{\includegraphics[width=\columnwidth]{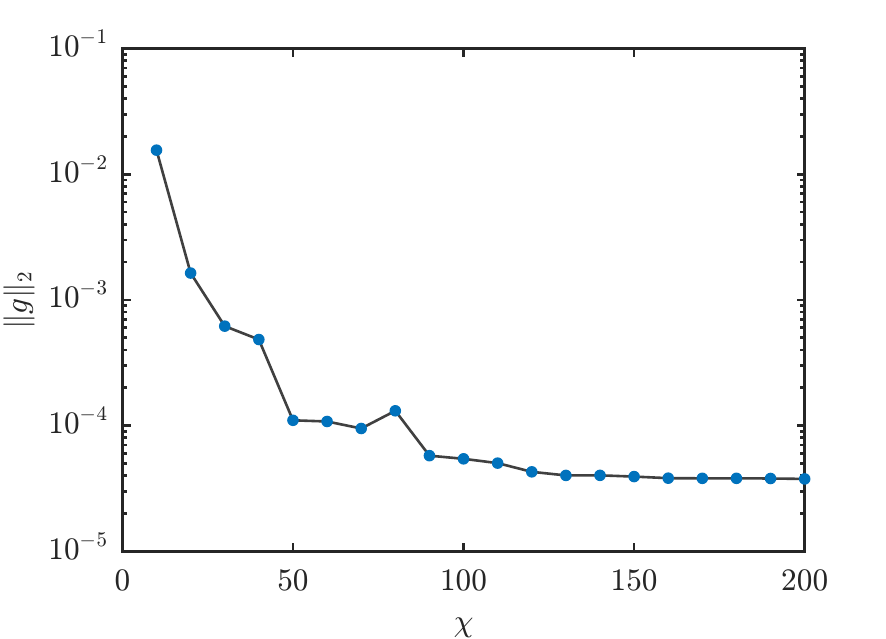}}
\caption{Details on the convergence of the optimization algorithm for the Ising model at $\lambda=3$. (Upper) The convergence of the norm of the gradient $\|g\|=\sqrt{g\dag g}$ (blue), the error in the energy (red) and the error in the magnetization (yellow), as a function of the iteration. The errors are computed as the relative error with respect to the last iteration. In this $D=2$ simulation the convergence criterion was $\|g\|\leq10^{-5}$, a value for which the two plotted observables have clearly converged. (Lower) The convergence of the norm of the gradient as a function of the bond dimension $\chi$ of the corner environment, at a particular iteration of the conjugate-gradient scheme for $D=3$ (close to convergence). This plot shows that large values of $\chi$ are needed to obtain a required tolerance on the norm of the gradient (in this case $\chi\approx100$).}
\label{fig:grad}
\end{figure}

 In Fig.~\ref{fig:grad} we provide some details on the convergence of the conjugate-gradient algorithm. In particular, we have found that rather high values of $\chi$ (the bond dimension of the corner environment) were needed to evaluate the gradient accurately close to convergence. Indeed, in the case of a strongly correlated PEPS, a lot of different terms contribute to the expression for the gradient. Close to convergence the gradient becomes a vector of small magnitude, which can only happen due to the subtle cancellations of a lot of different terms; consequently, finding the gradient accurately is bound to require a large value of $\chi$. Note that the large values of $\chi$ are only necessary close to convergence, so we grow $\chi$ throughout the optimization. We never impose the final value of $\chi$, because it is the correlations in the optimized PEPS that determine the $\chi$ needed to reach a certain tolerance on the norm of the gradient.

\par As a second application, we study two spin-1/2 Heisenberg models on the square lattice, defined by the Hamiltonian
\begin{equation*}
H_{\text{Heisenberg}} = \sum_{\braket{ij}} S^x_iS^x_j + S^y_iS^y_j + J_z S^z_iS^z_j .
\end{equation*}
The model has been of great theoretical and experimental interest, because of its paradigmatic long-range antiferromagnetic order \cite{Manousakis1991}. In particular, Heisenberg models have proven to be a hard case for the PEPS ansatz \cite{Bauer2009} because of the large quantum fluctuations around the antiferromagnetic ordering; as such, they provide a proper benchmark for our conjugate-gradient method.
\par
\begin{figure}
\subfigure{\includegraphics[width=0.49\columnwidth]{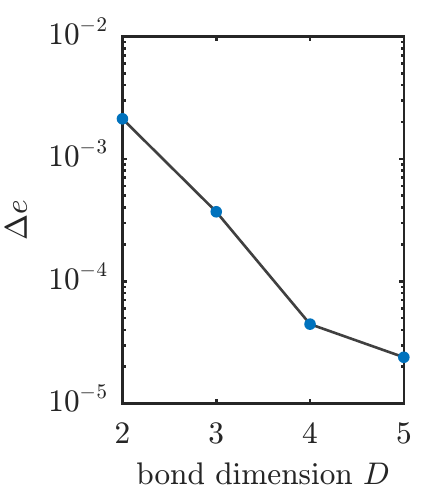}}
\subfigure{\includegraphics[width=0.49\columnwidth]{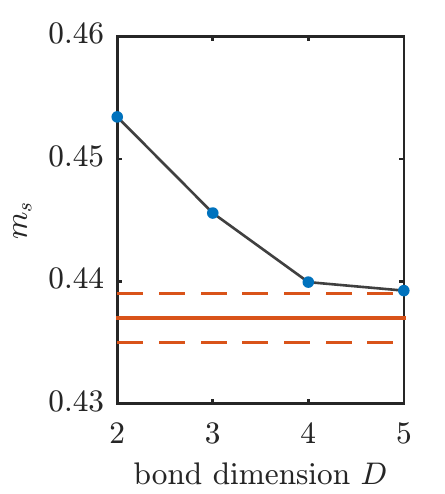}}
\caption{Results for the XY model ($J_z=0$), compared to the Monte Carlo results in Ref.~\onlinecite{Sandvik1999a}. (Left) The relative error $\Delta e= \left| (e_\text{var}-e_\text{MC})/e_\text{MC}\right|$ as a function of the bond dimension. (Right) The staggered magnetization as a function of the bond dimension; the red line is the Monte Carlo result with error bars.}
\label{fig:xy}
\end{figure}

\begin{figure}
\subfigure{\includegraphics[width=0.49\columnwidth]{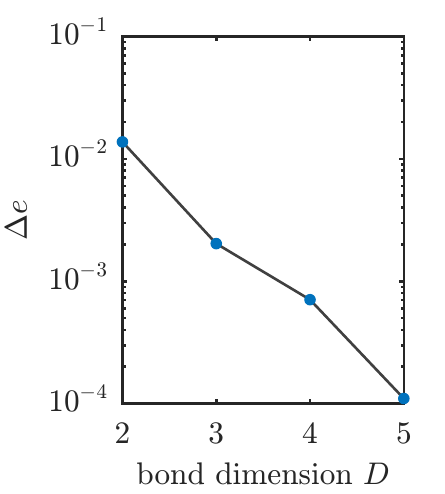}}
\subfigure{\includegraphics[width=0.49\columnwidth]{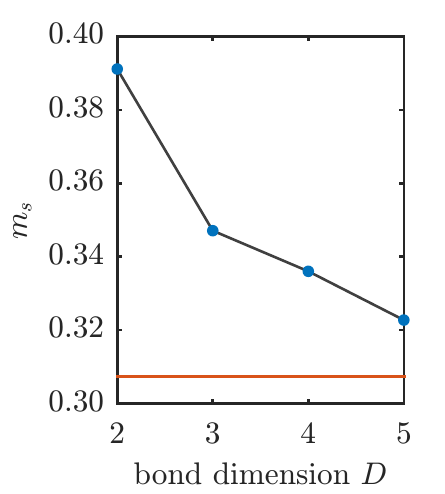}}
\caption{Results for the Heisenberg antiferromagnet ($J_z=1$), compared to the Monte Carlo results in Refs.~\onlinecite{Sandvik1997} and \onlinecite{Sandvik2010}. (Left) The relative error $\Delta e=\left| (e_\text{var}-e_\text{MC})/e_\text{MC}\right|$ as a function of the bond dimension. (Right) The staggered magnetization as a function of the bond dimension; the red line is the Monte Carlo result for which the error bars are too small to plot.}
\label{fig:heis}
\end{figure}

 In contrast to most PEPS implementations, we prefer to work with a single-site unit cell, so we perform a sublattice rotation in order to capture the staggered magnetic order in the ground state. Moreover, we impose rotational symmetry on the PEPS tensor $A$, so that our variational ground state is automatically invariant under rotations of the lattice. In Figs. \ref{fig:xy} and \ref{fig:heis} we have plotted the energy expectation value and staggered magnetization after convergence as a function of the bond dimension, for the XY model ($J_z=0$) and the isotropic Heisenberg model ($J_z=1$). Comparing with results from imaginary-time evolution \cite{Bauer2009, Phien2015}, we see that our variational method reaches considerably lower energies and order parameters at the same bond dimension.
\par
\begin{figure}
\includegraphics[width=\columnwidth]{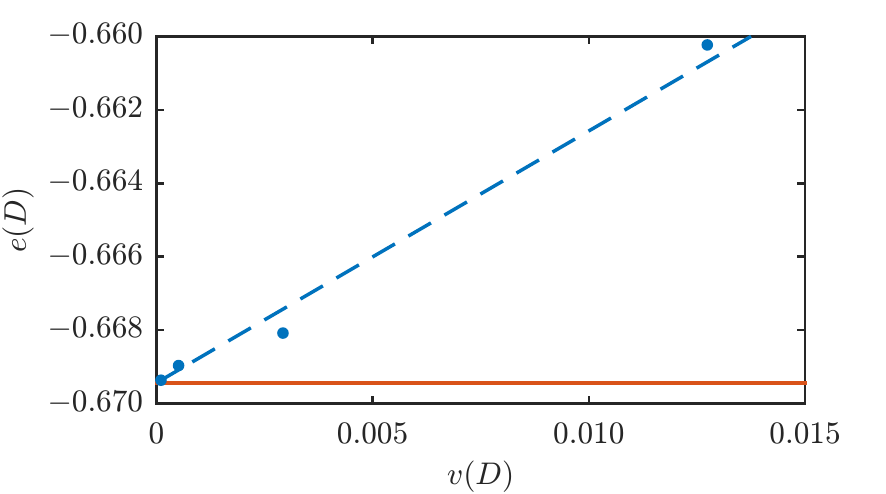}
\caption{The energy expectation value as a function of the variance per site for the isotropic ($J_z=1$) Heisenberg antiferromagnet, for four values of the bond dimension ($D=2\rightarrow5$). The red line represents the Monte Carlo result for the ground-state energy \cite{Sandvik1997}. A linear extrapolation with respect to the two best points ($D=4,5$) gives an energy with a relative error of $\Delta e\approx 4.7\times10^{-5}$. The striped line is drawn between the exact MC result and the $D=5$ point and serves only as a guide to the eye.}
\label{fig:var}
\end{figure}

 In addition we also compute the variance of these PEPS variational states, in order to get an idea of how well they approximate the true ground state. The result for the isotropic Heisenberg model is plotted in Fig.~\ref{fig:var}. We observe the expected linear behavior \cite{Sorella2001} to some extent, and a zero-variance extrapolation based on the two last points ($D=4,5$) improves the estimate of the energy by a factor of two. Better zero-variance extrapolations should be possible at higher bond dimensions for which the linear behavior is expected to be stronger.
\par
\begin{figure}
\includegraphics[width=\columnwidth]{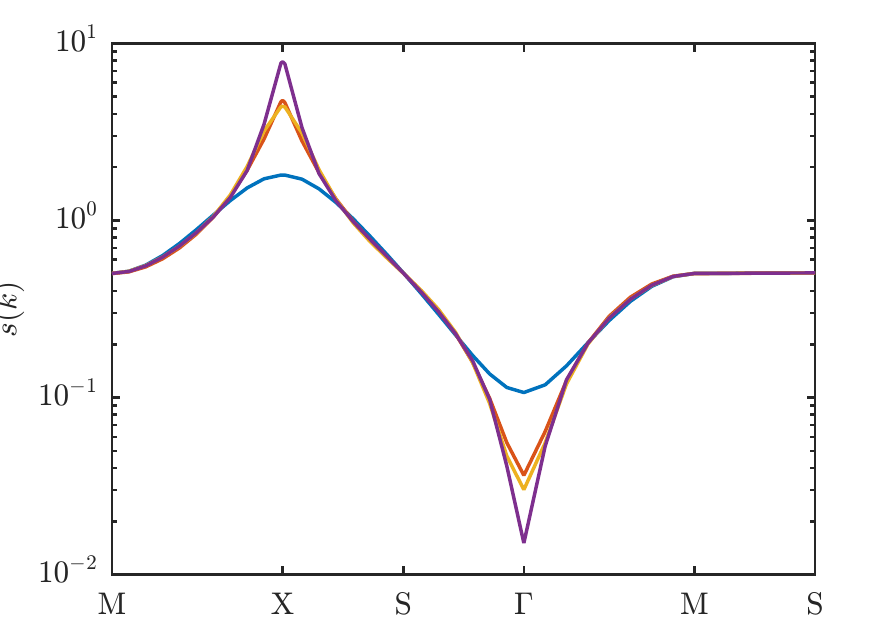}
\caption{Structure factor $s(\vec{q})$ of the isotropic ($J_z=1$) Heisenberg antiferromagnet along a path through the Brillouin zone for optimized PEPS states with bond dimensions $D=2$ (blue), $D=3$ (red), $D=4$ (orange), and $D=5$ (purple), in agreement with the results in Refs. \onlinecite{Zheng2005} and \onlinecite{Luscher2009}. The divergence around the X point and the zero around the $\Gamma$ point are better reproduced as $D$ increases, although the improvement as a function of $D$ seems not to be smooth.}
\label{fig:sf}
\end{figure}

 Another quantity that is within reach of our PEPS framework is the static structure factor, a central quantity for detecting the order in the ground state, and of direct experimental relevance. It is defined as
\begin{equation*}
s(\vec{q}) = \frac{1}{|\mathcal{L}|} \sum_{i,j\in\mathcal{L}} \e^{i\vec{q}\cdot(\vec{n}_i-\vec{n}_j)} \braket{\vec{S}_i \cdot \vec{S}_j}_c, 
\end{equation*}
where only the connected part of the correlation function is taken into account. The disconnected part will give a $\delta$-peak at $\vec{q}=(\pi,\pi)$ (the $X$ point), corresponding to the staggered-magnetization order parameter. The strong fluctuations around this point will give an additional $1/q$ divergence, with $q$ the distance from the $X$ point \cite{Zheng2005}. The structure factor becomes zero at $\vec{q}=(0,0)$, because the ground state is in a singlet state. In Fig.~\ref{fig:sf} we observe that the regular parts of the structure factor are perfectly reproduced, even at low bond dimensions, whereas the divergences can only be accurately captured by observing the behavior as a function of the bond dimension.


\section{Conclusions}
\label{sec:concl}

In conclusion, we have presented an algorithm for numerically optimizing the PEPS ansatz for ground-state approximations. The algorithm is based on the efficient evaluation of the energy gradient, and is a direct implementation of the variational principle with a clear convergence criterion. Starting from a random PEPS tensor, it allows us to find a variational minimum for a given bond dimension. 
\par As such, our approach is complementary to any other PEPS algorithm. In fact, our variational search systematically finds lower energies than algorithms based on imaginary-time evolution and local truncations. This observation is consistent with the recent results in Ref.~\onlinecite{Corboz2016a}, where an alternative variational algorithm was proposed. This confirms our belief that a variational approach will be crucial in the future for capturing, e.g., phase transitions in two-dimensional lattice systems. 
\par Our approach has the additional advantage that global symmetries can be exploited easily, which should lead to more efficient simulations \cite{Bauer2011a}. Moreover, the implementation of symmetries will prove crucial for simulating systems with topological order, which can be imposed as a matrix product operator symmetry on the virtual level of the PEPS tensor \cite{Bultinck2015}. Finally, our approach straightforwardly allows us to consider reduced PEPS parametrizations by confining our optimization scheme to a certain PEPS subclass \cite{Poilblanc2013a,prepMichael}.
\par In addition, some of the methods that we have presented in this paper could be applicable to the variational optimization of PEPS on finite lattices \cite{Verstraete2004b, Murg2007, Murg2009, Pizorn2010, Wang2011, Lubasch2014, Jiang2015} as well. With finite PEPS simulations, the straightforward approach of optimizing the different PEPS tensors sequentially is severely hampered by the bad conditioning of the normalization matrix. In particular, because the energy and normalization matrix require different effective environments, the regularization of this bad condition number is not well defined. With a finite-lattice version of the corner environments, however, we could use the same effective environment for computing the energy and normalization, allowing a consistent regularization of both the energy and normalization matrix. This should lead to efficient variational optimization methods for finite PEPS as well.
\par Our framework has allowed us to compute the structure factor, which is of direct experimental relevance, and the energy variance, which provides an unbiased measure of the variational error of the PEPS ansatz. Although the variance extrapolations seem to be not straightforwardly implementable, this should contribute to better energy bond dimension extrapolations in the future. For systems with a number of competing ground states such as the Hubbard model \cite{Corboz2016}, this extrapolation will be of crucial importance.
\par From the perspective of numerical optimization, a conjugate-gradient search is only a first step to more advanced schemes such as Newton or quasi-Newton methods. The Hessian of the energy functional is crucial in these optimization schemes, the evaluation of which is straightforward with our effective environment. Alternatively, the non trivial geometric structure of the PEPS manifold can be taken into account in the optimization \cite{Absil2008}. Also, imposing a certain gauge fixing on the PEPS tensor might render the optimization more efficient.
\par Finally, it seems that tangent-space methods that have proven successful in the context of matrix product states \cite{Haegeman2013b} are now within reach for PEPS simulations for generic two-dimensional quantum spin models. In particular, this paper opens up the prospect of simulating real-time evolution according to the time-dependent variational principle \cite{Haegeman2011d} and/or computing the low-energy spectrum on top of a generic PEPS with the quasiparticle excitation ansatz \cite{Haegeman2012a, Vanderstraeten2015a}.
\par This research was supported by the Research Foundation Flanders (L.V. \& J.H.), by the Delta-ITP [an NWO program funded by the Dutch OCW] (P.C.), and by the Austrian FWF SFB through Grants FoQuS and ViCoM and the European Grants SIQS and QUTE (F.V.).

\bibliography{bibliography}

\end{document}